\documentclass[aps,prd,showpacs,preprintnumbers]{revtex4}
\usepackage{amsmath,amssymb}
\usepackage{graphicx}
\usepackage{subfigure}
\usepackage{color}

\newcommand\bef{\begin{figure}}
\newcommand\eef[1]{\label{fig.#1}\end{figure}}
\newcommand\beq{\begin{equation}}
\newcommand\eeq[1]{\label{eq.#1}\end{equation}}
\newcommand\beqa{\begin{eqnarray}}
\newcommand\eeqa[1]{\label{eq.#1}\end{eqnarray}}
\newcommand\bet{\begin{table}}
\newcommand\eet[1]{\label{tbl.#1}\end{table}}
\newcommand\betb{\begin{center}\begin{tabular}}
\newcommand\eetb{\end{tabular}\end{center}}
\newcommand\beit{\begin{itemize}}
\newcommand\eeit{\end{itemize}}

\newcommand\fig[1]{Fig.\ \ref{fig.#1}}
\newcommand\eq[1]{Eq.\ (\ref{eq.#1})}

\newcommand{\erfc}{\Phi_c}
\newcommand{\cf}{C_f}
\newcommand{\chil}{\chi_{\scriptscriptstyle P}}

\newcommand{\tav}[1]{\langle  #1 \rangle_{\scriptscriptstyle T}}
\newcommand{\tc}{T_c}

\newcommand{\tT}{\sqrt{t} T}
\newcommand{\prent}{P_{\rm ren}(T, t)}
\newcommand{\pren}{P_{\rm ren}(T)}
\newcommand{\pms}{P^{\overline{\scriptscriptstyle{MS}}}(T)}
\newcommand{\gms}{g^2_{\overline{\scriptscriptstyle{MS}}}(\mu=1/\sqrt{8 t})}
\newcommand{\lms}{\Lambda^{n_f=0}_{\overline{\scriptscriptstyle{MS}}}}

\newcommand{\Eav}{\langle E(T, t) \rangle}
\newcommand{\Mav}{\langle M(T, t) \rangle}
\newcommand{\EMav}{\langle E-M \rangle (T,t)}
\newcommand{\Gsq}{\langle G^2 \rangle }
\newcommand{\GsqT}{\langle G^2 \rangle_T }
\newcommand{\GesqT}{\langle G_{\scriptstyle E}^2 \rangle_T }
\newcommand{\GmsqT}{\langle G_{\scriptstyle M}^2 \rangle_T }
\newcommand{\Gbar}{\langle \underline{G^2} \rangle_T}
\newcommand{\Ebar}{\langle \underline{G_{\scriptstyle E}^2} \rangle_T}
\newcommand{\Mbar}{\langle \underline{G_{\scriptstyle M}^2} \rangle_T}
\newcommand{\me}{m_{\scriptscriptstyle E}}
\newcommand{\ebf}{\bf E}

\newcommand\tr{\mathrm{Tr}\;}
\newcommand{\GeV}{\mathrm{GeV}}

\newcommand\epj{{\sl Eur.\ Phys.\ J.\/}\ }
\newcommand\jhep{{\sl J.\ H.\ E.\ P.\/}\ }
\newcommand\np{{\sl Nucl.\ Phys.\/}\ }

\newcommand\pr{{\sl Phys.\ Rev.\/}\ }

\newcommand\ptep{{\sl Prog.\ Theor.\ Exp.\ Phys.\/}\ }

\begin{document}
\title{Using Wilson flow to study the SU(3) deconfinement transition}
\author{Saumen Datta}
\email{saumen@theory.tifr.res.in}
\affiliation{Department of Theoretical Physics, Tata Institute of Fundamental
         Research,\\ Homi Bhabha Road, Mumbai 400005, India.}
\author{Sourendu\ \surname{Gupta}}
\email{sgupta@theory.tifr.res.in}
\affiliation{Department of Theoretical Physics, Tata Institute of Fundamental
         Research,\\ Homi Bhabha Road, Mumbai 400005, India.}
\author{Andrew\ \surname{Lytle}}
\email{andrew.lytle@glasgow.ac.uk}
\affiliation{SUPA, School of Physics and Astronomy, University of Glasgow,
         Glasgow G12 8QQ, UK.}
\begin{abstract}
We explore the use of Wilson flow to study the deconfinement transition in
SU(3) gauge theory. We use the flowed Polyakov loop as a renormalized order
parameter for the transition, and use it to renormalize the Polyakov loop. 
We also study the flow properties of the electric and magnetic gluon 
condensates, and demonstrate that the difference of the flowed operators 
shows rapid change across the transition point. 
\end{abstract}

\preprint{TIFR/TH/15-37}

\maketitle

\section{Introduction}
\label{sec.intro}
Wilson flow is a powerful new technique for the study of non-Abelian
gauge theories \cite{main, neuberger}.  It has been used for
setting the scale in lattice computations
\cite{main,bmw,staggered,scale,milc}. 
It can also be
applied in the construction of renormalized composite operators, like the
energy-momentum tensor \cite{suzuki,dpr} and fermion bilinears
\cite{chiral,endo,luscher}. 
 One example of the use of operators renormalized this way is 
the recent attempt to extract the renormalized pressure and energy
density at finite temperature, $T$, in SU(3) gauge theory
\cite{flowqcd}. In this paper we use Wilson flow to create an order
parameter for the finite temperature transition in the SU(3) pure
gauge theory, and to examine gluon condensates for $T > 0$.

The Wilson flow equation \\ 
\beq 
\frac{d U_\mu(x, t)}{dt} \ = \ - \, \partial_{x, \mu} S[U] \,  
\cdot \, U_\mu(x, t), 
\eeq{formflow} 
produces a smeared configuration, $U_\mu(x, t)$, at any ``flow time''
t, given the initial condition $U_\mu(x, 0) = U_\mu(x)$. Here
$U_\mu(x)$ is the bare link ($x$ denotes a point in the 4-d Euclidean
space-time lattice, and $\mu$ denotes one of the 4 directions),
$\frac{\textstyle 1}{\textstyle g_0^2} S[U]$ is the action, and the
derivative is a Hermitian traceless matrix. In this paper we will use
the Wilson action, and our convention will be that
\beq 
S[U] = \sum_p {\rm Re} \ \tr \left[ 1 - U(p) \right]. 
\eeq{action} 
Here the plaquette operator, $U(p)$, is the ordered product of link
matrices around a plaquette, and the sum is over all oriented
plaquettes; $\partial_{x, \mu} \, S[U]$ in \eq{formflow} is the
traceless Hermitian matrix constructed from $U(p)$, the plaquette $p$
containing the link $(x, \mu)$.

Since the flow defined by \eq{formflow} is diffusive, the smeared link
operator has size which is proportional to $\sqrt{t}$. If one could choose to
work at a flow time $t$, fixed in physical units while changing the
lattice spacing, then the fat-link operators $U_\mu(x, t)$ would all be
evolved to the same physical scale, and one would be able to construct
renormalized composite operators from them. We explore such a
construction for the Wilson line here.

A common way to define the scale is through the gluon condensate \cite{main}: 
\beq
{\mathcal E}(t) = t^2 {\bf E}(t), \qquad {\rm where} \qquad 
{\bf E}(t) = - \frac{1}{2} \overline{\tr G_{\mu \nu}(x, t) G_{\mu \nu}(x, t)}.
\eeq{cond}
$G_{\mu \nu}$ is the lattice version of the field strength tensor,
and an average over the 4-volume of the lattice is denoted by
the bar. One selects a value of $c$ and solves the equation
\beq
\langle {\mathcal E}(t) \rangle = c
\eeq{t0}
for $t$. We will denote such choices of $t$ by $t_c$. The specific choice 
$c$ = 0.3 defines the flow time $t_{0.3}$, which is commonly called $t_0$ 
\cite{main}. Another suggestion has been to use a derivative of 
$\mathcal{E}(t)$ \cite{bmw}. Systematics of these scale setting schemes 
have been studied in detail \cite{scale,staggered,milc}. 

$\langle \mathcal{E} \rangle$ has widely been used for scale setting
purposes in lattice QCD. In perturbation theory
\beq
\langle \mathcal{E} \rangle \ = \ \frac{3}{16 \pi^2} \; 
g^2 \left( 1 + c_1 g^2 + \mathcal{O} (g^4) \right)
\eeq{condpert}
where, for pure gauge SU(3), 
$c_1 = \frac{\textstyle 1.0978}{\textstyle 4 \pi}$ when $g^2 \equiv
g^2_{\overline{\scriptscriptstyle{MS}}} (\mu=1/\sqrt{8 t})$
\cite{main}.  One can also use $ \langle \mathcal{E} \rangle$ to
define a new coupling scheme, 
\beq
g^2_{\rm flow}(t) \ = \ \frac{16 \pi^2}{3} \langle \mathcal{E} \rangle.
\eeq{flowcoupling}

In this paper we investigate the flow of two quantities which are very
sensitive to the deconfinement transition. First, we look at the Polyakov loop,
which is the order parameter for the deconfinement transition, but is
highly singular as one takes the continuum limit. We discuss in the
next section the use of flow to construct a continuum order
parameter. Renormalization of the Polyakov loop using Wilson flow has also
been considered in Ref.~\cite{petreczky}, which considered Polyakov loops
in various representations, though we take a somewhat different
approach to renormalizing them than what was done there.
In the following section, we discuss the flow-time behavior
of the gluon condensate and related observables. The gluon condensate
is related to the nonperturbative nature of the QCD vacuum. As one
crosses the deconfinement temperature $\tc$, the gluon condensate 
starts to melt. Also the electric
and magnetic components of the gluon condensate show different
temperature dependences. We will see that flow enhances the
sensitivity of the gluon condensate to the onset of the transition.

In order to reduce the dependence of observables on the ultraviolet scale 
$1/a$, one should choose $c$ such that $a \ll \sqrt{t}$.
At finite temperature, $T$, there is also an infrared scale proportional 
to $T$, and one should ideally choose
\beq
T \ll \frac{1}{\sqrt{t}} \ll \frac{1}{a}. 
\eeq{range}
Since one also has $T N_t = 1/a$, one sees that the hierarchies imply
$1/N_t \ll \sqrt{t}T \ll 1$. With current day lattice $N_t \le 16$, so
the practical interpretation of ``much less than'' is no better than a
factor of 4. As a practical example, when one chooses $c = 0.3$, so
that the flow scale is $t_0$, then $T_c \sqrt{t_0} \approx $ 0.25 in
pure SU(3) gauge theory \cite{tct0}.  Also, in most computations
today, $\sqrt{t_0}/a \lesssim 2$.  The process of scanning in $T$
while keeping $1/\sqrt{t}$ fixed (by fixing $c$) means that the
hierarchy in \eq{range} can be preserved only for $1/(N_t \sqrt{t})
\ll T \ll 1/\sqrt{t}$. The suggestion in Ref. \cite{flowqcd}, that one
could keep $\sqrt{t}T = b$ fixed as one changes $T$ obviously has the
limits $1/N_t \ll b \ll 1$.  We study these questions here as part of
our study of the renormalized Polyakov loop and gluon condensates at
finite temperature.

\section{Polyakov loop}
\label{sec.poly}

The deconfinement transition is associated with the breaking of the
$Z_3$ center symmetry for SU(3) gauge theory. The Polyakov loop,
\beq 
L(T, a) \ = \ \frac{1}{3 V} \ \sum_{\bf x} \ \tr
\prod_{x_4=1}^{N_t} U_t (x)
\eeq{polloop} 
transforms nontrivially under the $Z_3$ symmetry and acts as an order
parameter for the transition. Here $x=({\bf x}, x_4)$ are the
coordinates of the lattice sites, $U_4(x)$ are the link elements at
site $x$ in the Euclidean time direction, $a$ is the lattice spacing,
$N_t$ is the number of sites in the Euclidean time direction, the
temperature, $T = 1/(a N_t)$, and the volume $V = N_s^3$ where $N_s$
is the number of sites in the spatial directions.

$\tav{L(T,a)} = 0$ for temperature $T<T_c$, where the center symmetry
is unbroken. Here $\langle \cdot \cdot \rangle_{\scriptscriptstyle T}$
denotes thermal averaging. For $T > T_c$ the $Z_3$ symmetry is
spontaneously broken, and $\tav{L(T, a)} $ becomes nonzero. At finite
volume, tunnelling between the $Z_3$ vacua make $\tav{L(T, a)} \to 0$
even in deconfiment phase; so we follow the standard practice of
studying
\beq
P(T, a) \ = \ \langle |L(T, a)| \rangle_{\scriptscriptstyle T}.
\eeq{ploop}
$P$ is nonzero below $\tc$, 
$P \vert_{\scriptscriptstyle T < \tc} \sim \frac{\textstyle
  1}{\textstyle V}$.
   
The bare Polyakov loop, as defined in \eq{ploop}, depends strongly on 
the lattice spacing $a$ \cite{polyakov}: \\
\beq
P(T, a) \ = \ e^{-f(g^2(a))/aT} \ P_{\rm ren}(T).
\eeq{polren}
Therefore $P(T, a) \to 0$ as $a \to 0$ and needs to be renormalized.
Various techniques for renormalizing the Polyakov loop have been
proposed in the literature \cite{renpol,renpol1}. The renormalized
Polyakov loop has also been calculated to next-to-leading-order in 
perturbation theory
\cite{mikko}; in the $\overline{\rm MS}$ scheme,
\beq
\log \pms = 1 + \frac{g^2 \cf \me}{8 \pi T} 
+ \frac{3 g^4 \cf}{16 \pi^2} \left( \log \frac{\me}{T} + \frac{1}{4} \right) 
+ .....
\eeq{msbar}
where $g^2$ is the coupling in $\overline{\rm MS}$ scheme at a scale
$\sim 4 T$ and $m_E$ is the electric screening mass. For SU(3) gauge
theory, $m_E = g T$ in leading order of perturbation theory.

\subsection{Flowed Polyakov Loop}
\label{sec.flowpoly}
Wilson flow can be used to define an order parameter that is only mildly
dependent on the lattice spacing $a$, and has a finite continuum limit: if
we flow to a physical scale $t$, and define a Polyakov loop, $P(T, t,
a)$ through \eq{polloop} with the links replaced by flowed links, then 
$P(T, t, a) = P(T, t) + \mathcal{O}(a^2/t)$. Since the  
Wilson flow preserves center symmetry, the flowed Polyakov loop 
$P(T, t, a)$ acts as an order parameter for the deconfinement
transition.

As discussed in Sec. \ref{sec.intro}, if we flow the fields to time $t$, 
operators constructed out of the flowed fields are smeared to a 
radius $\sim \ \sqrt{8 t}$. So we expect finite $a$ corrections to be small 
for  $\sqrt{8 t} \gg a$. On the other hand, for thermal physics we require 
the smearing radius $\sqrt{8 t} \ll 1/T$. A window of $t$ satisfying
both conditions can be obtained for the kind of lattices commonly used
for finite temperature physics \cite{flowqcd}. 

In \fig{polflow}, we show the flowed Polyakov loop for three different
lattice spacings, corresponding to $N_t$ =6, 8 and 10, respectively,
at two different temperatures. At $t=0$ we see the strong $a$
dependence indicated by \eq{polren}. We see that this divergence is
removed at fairly early flow times, $\tT \simeq 0.05$.  The remaining
finite $a$ corrections are suppressed when the flow time increases to
$\tT \simeq 0.16$. If one does not include the $N_t$ = 6 data, then
the figure shows that this happens at $\tT \simeq 0.12$.  These flow
times saturate the lower bound $b > 1/N_t$, and correspond to
$\sqrt{t}/a \simeq 1$ on the respective lattices. This is good as a
practical matter, since it implies that the ``much less than'' in
\eq{range} can be replaced by ``less than''.

\bef
\includegraphics[scale=0.7]{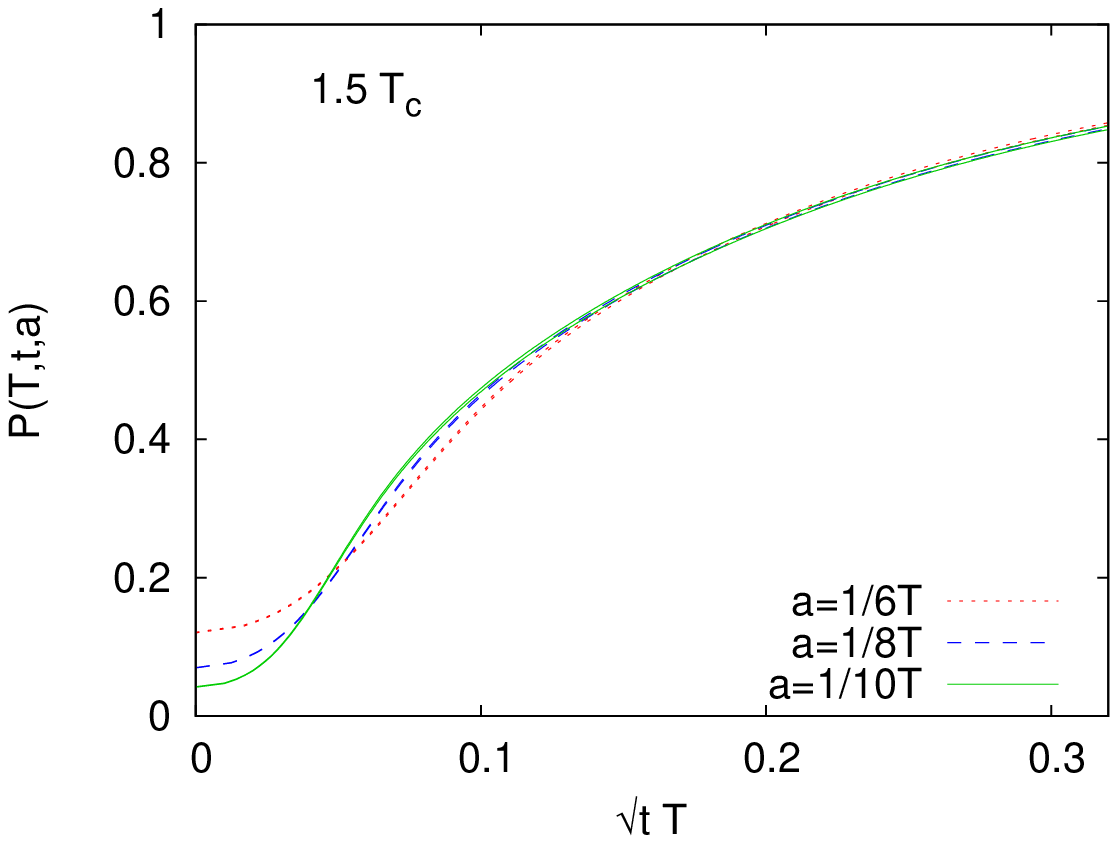}
\includegraphics[scale=0.7]{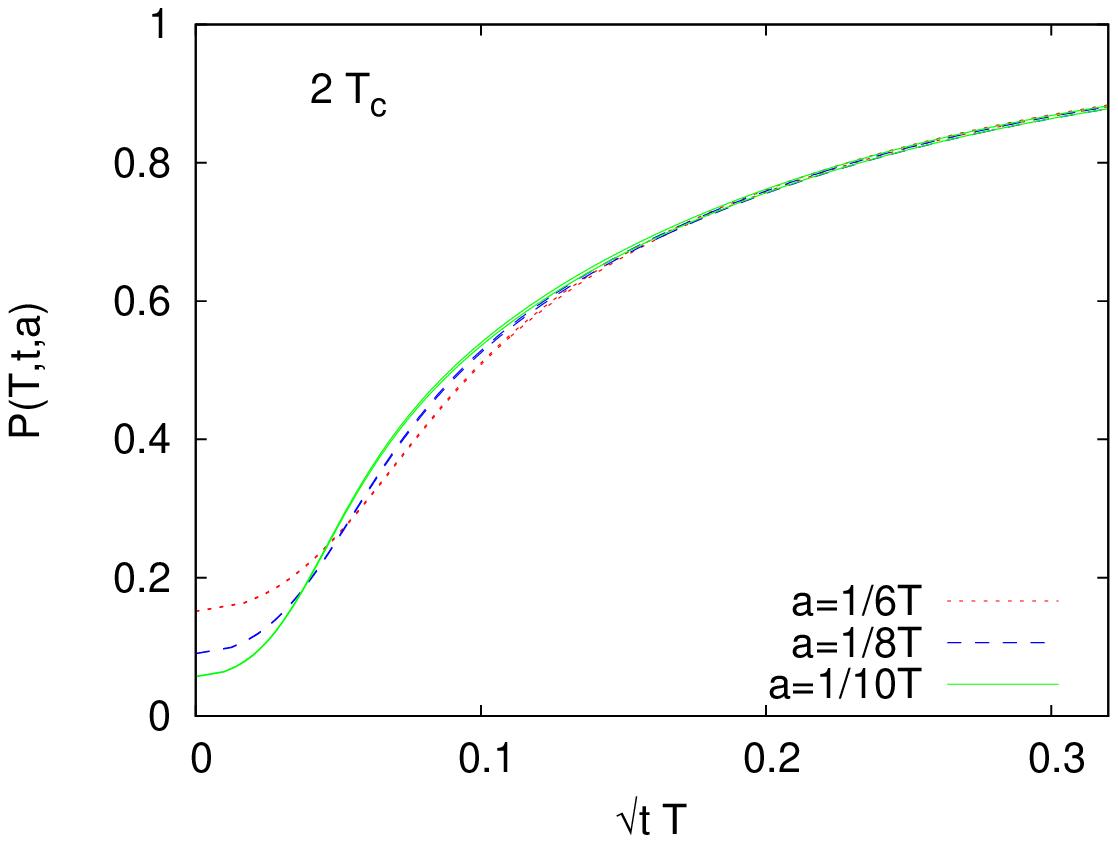}
\caption{Flowed Polyakov loop at 1.5 $T_c$ (left) and 2 $T_c$ (right). The 
thickness of the lines represent the $1 \sigma$ bands.}
\eef{polflow}

In \fig{flowT} we explore the temperature dependence of the flowed Polyakov 
loop, $P(T, t, a)$, at three different lattice spacings, where the flow time 
is fixed to $t_{0.15}$ (left) and to $t=1/(5 T)^2$ (right). As discussed in 
Appendix \ref{sec.detail} the critical temperature is obtained from the 
peak of the susceptibility of the bare Polyakov loop, while $T/T_c$ is
obtained using the flow scale. Since flowing to a fixed length 
scale like $t_{0.15}$ interferes with thermal physics at sufficiently high 
temperatures, in the left panel above we had to stop at a temperature 
$ < 1/\sqrt{8 t_{0.15}}$.  We note that 
$\sqrt{t}/a$ is large enough that the difference between the flowed 
loops at different $a$ are too small to be seen, for both these choices 
of $t$. Whenever our choice of flow time allows us to do this, we will 
suppress the argument $a$ and refer to $P(T, t)$. 
 
This $a$-independent flowed Polyakov loop is sufficient to measure the
continuum deconfinement transition in pure gauge theory. In the lower
panels of \fig{flowT}, we show the susceptibility density \\
\beq
\chil (T, t) = \tav{|P(T, t)|^2} - \tav{|P(T, t)|}^2 .
\eeq{susc}
Since SU(3) gauge theory has a first order transition, 
$\chil (T, t)$
is expected to show a peak at $T_c$, just like the susceptibility for
the non-flowed loop. Unlike the latter, however, the flowed
susceptibility peak height does not change with $a$. The
susceptibility $V \chil$ is known to scale like volume at $T_c$;
since this volume scaling is caused by the two-peak nature of $P$ at
the transition point, one expects a similar scaling to hold here. We
do not explicitly check this volume dependence here.

\bef
\includegraphics[scale=0.7]{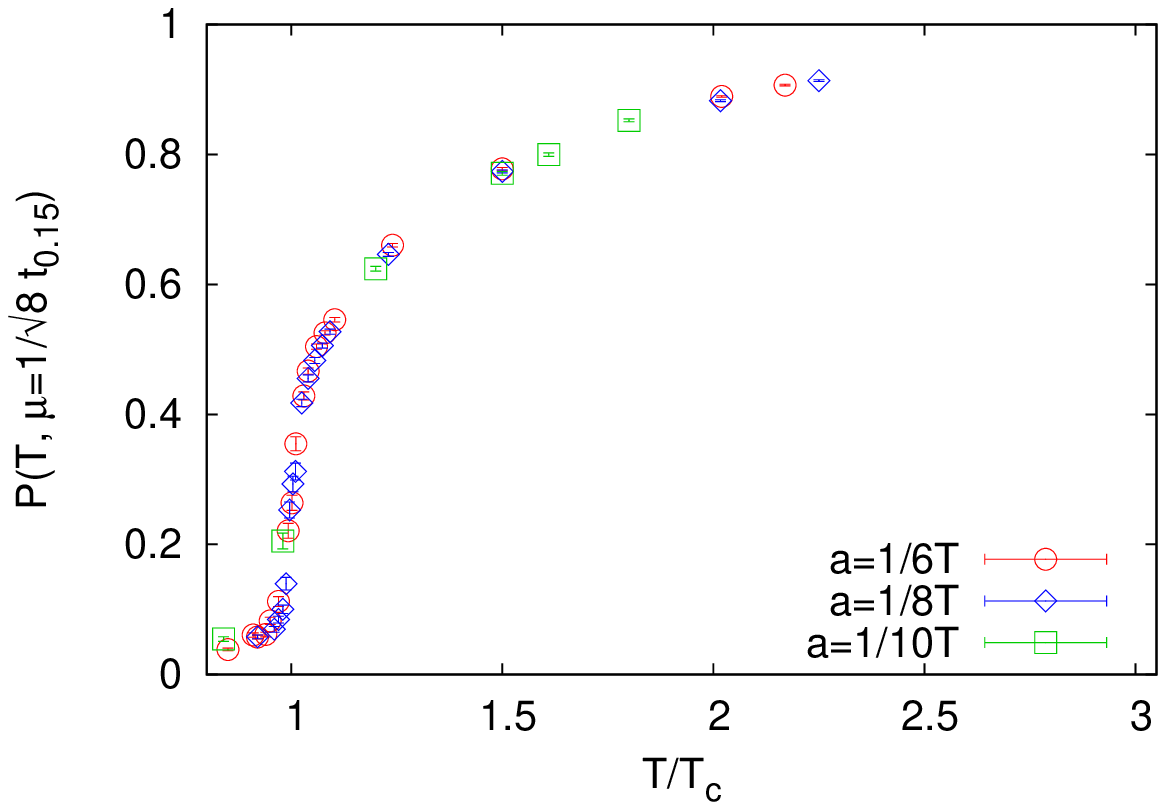}
\includegraphics[scale=0.7]{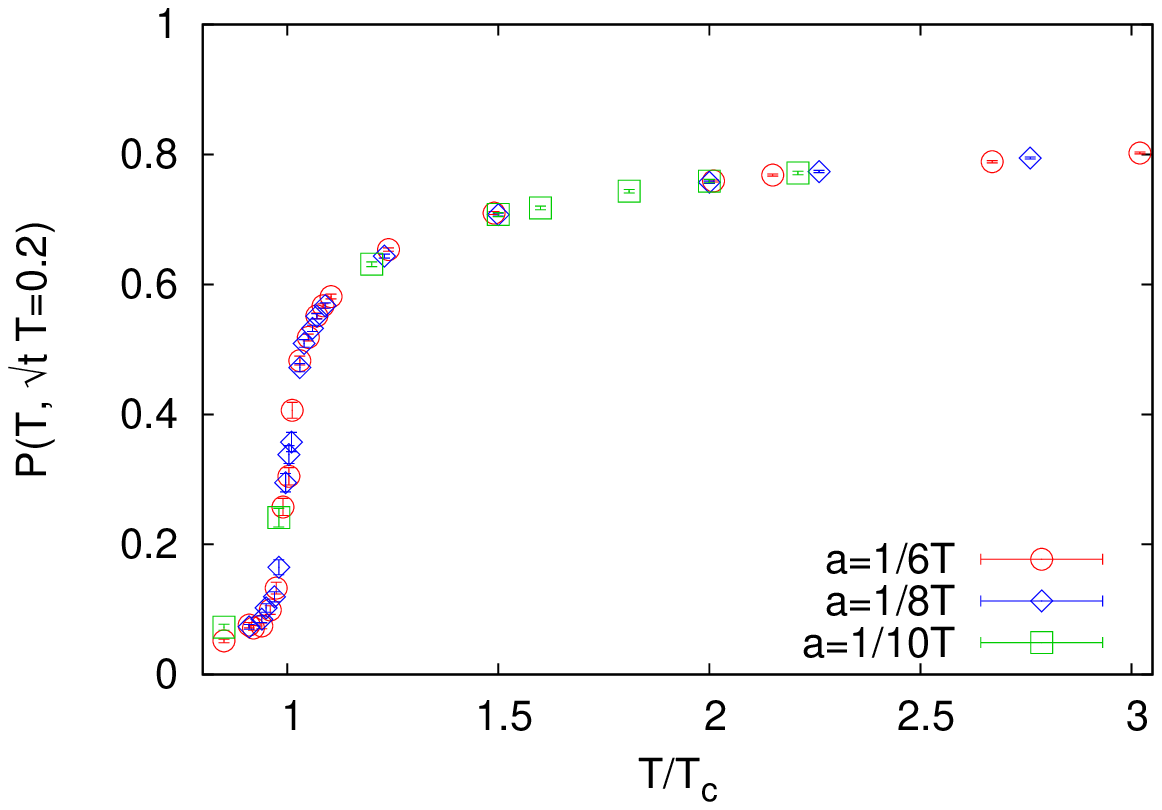}
\includegraphics[scale=0.7]{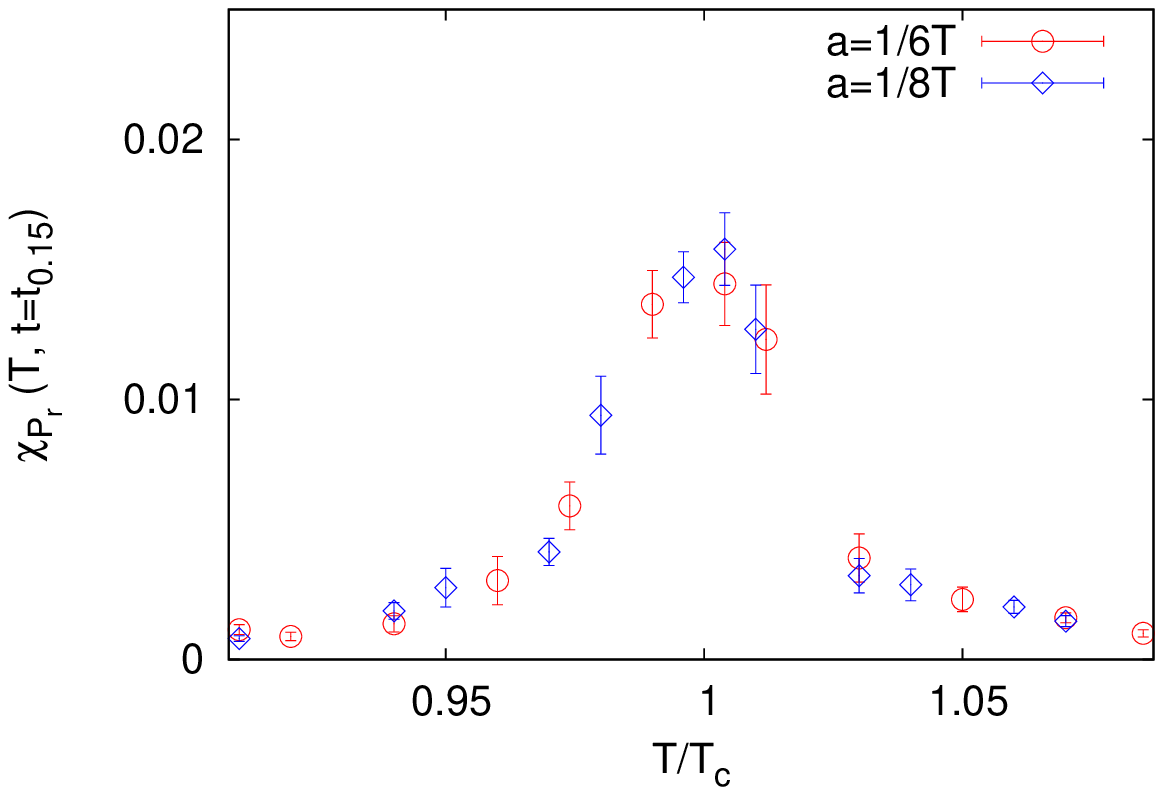}
\includegraphics[scale=0.7]{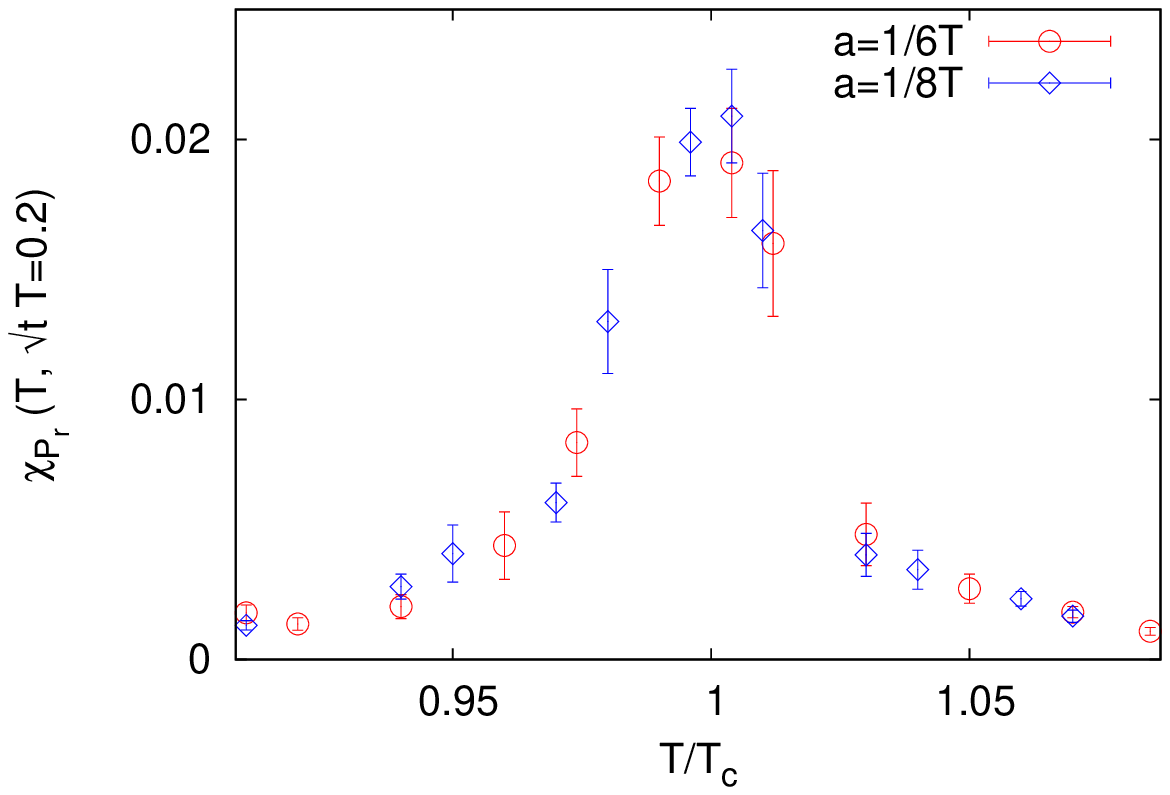}
\caption{The flowed Polyakov loop, $P(T, t, a)$, and its susceptibility 
  density, $\chil (T, t, a)$, at fixed $t_0$ (left)
  and at fixed $\sqrt{t}T$ (right). The dependence on $a$ is too small
  to be extracted from these measurements. The thermal transition is
  identified correctly by both measures as one can see from the fact
  that the peak of the susceptibilities coincides with
  the $T_c$ measurement using the bare Polyakov loop.}
\eef{flowT}

The symmetries of the Polyakov loop decide which screening masses can
be seen in their correlations. Since the symmetries of the bare and
flowed Polyakov loops are the same, they would give the same screening
masses. The Polyakov loop correlations are
sometimes used to determine the free energy of an infinitely heavy
colour source placed in the gluonic medium. Determining this would
require a renormalized Polyakov loop,whose extraction from data we
turn to next.

\subsection{Renormalized Polyakov loop}
\label{sec.polren}

For gauge links flowed to a (sufficiently large) flow time $t$,
fluctuations at scale $\gg 1/\sqrt{t}$ are strongly suppressed and 
the effective ultraviolet cutoff is $\sim 1/\sqrt{8 t}$ \cite{main}. 
Therefore, similar to \eq{polren} we can write \\
\beq
P(T, t) = e^{-\frac{R \left( g^2(t) \right) }{\sqrt{t} T}} \ \pren
\eeq{prent}
where $g^2(t)$ is the coupling evaluated at a scale $\mu \sim 1/\sqrt{8 t}$.
In leading order in $g$ (see Appendix \ref{pol.lo}), \\
\beq
R \left(g^2(t) \right) = 
\frac{1}{3 \pi^2} \frac{\sqrt{\pi}}{\sqrt{8}} \ g^2(t) \ 
\left( 1 + \mathcal{O}(g^2) \right) .
\eeq{expr}
Following standard arguments \cite{polyakov, dotsenko} we expect that 
$\exp(R/\sqrt{t} T) P(T, t) \equiv \prent$  
is a function of temperature modulo $\mathcal{O}(\sqrt{t} T)$ corrections, 
and has a finite limit as $t \to 0$. 

\bef
\includegraphics[scale=0.7]{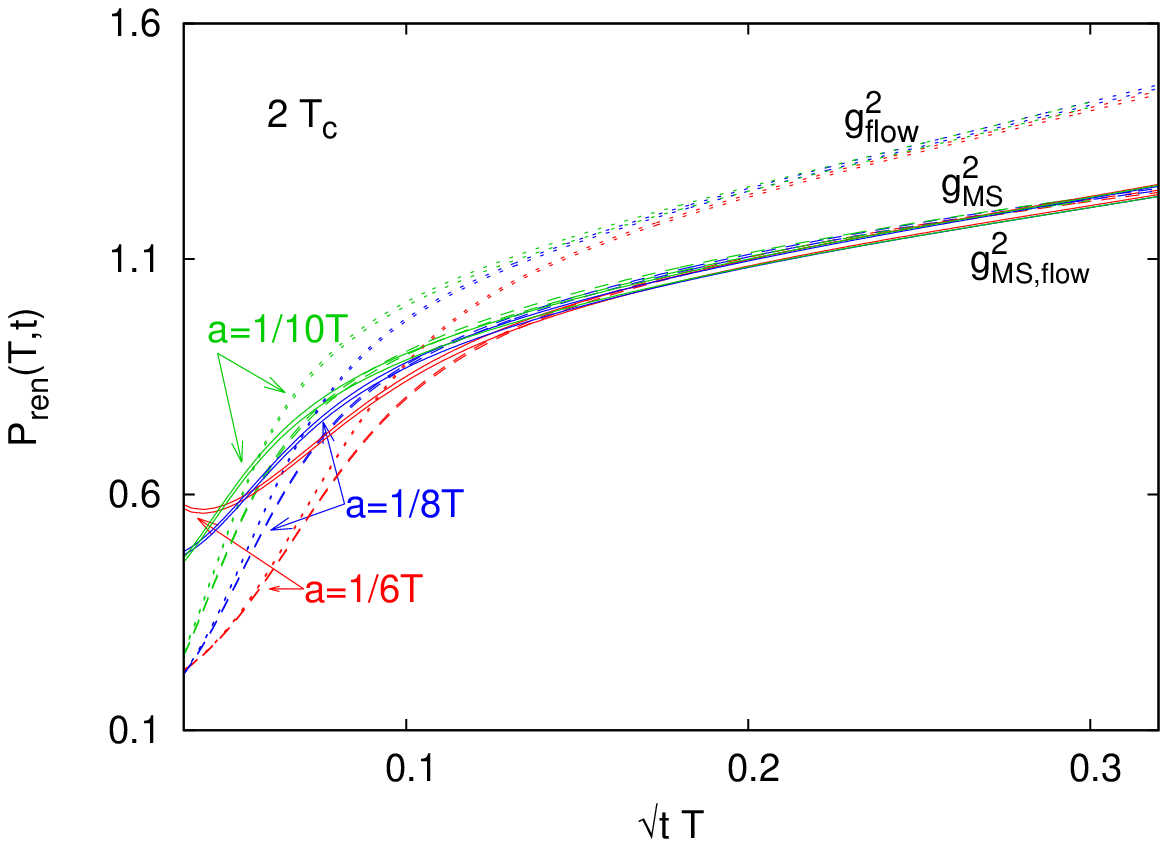}
\includegraphics[scale=0.7]{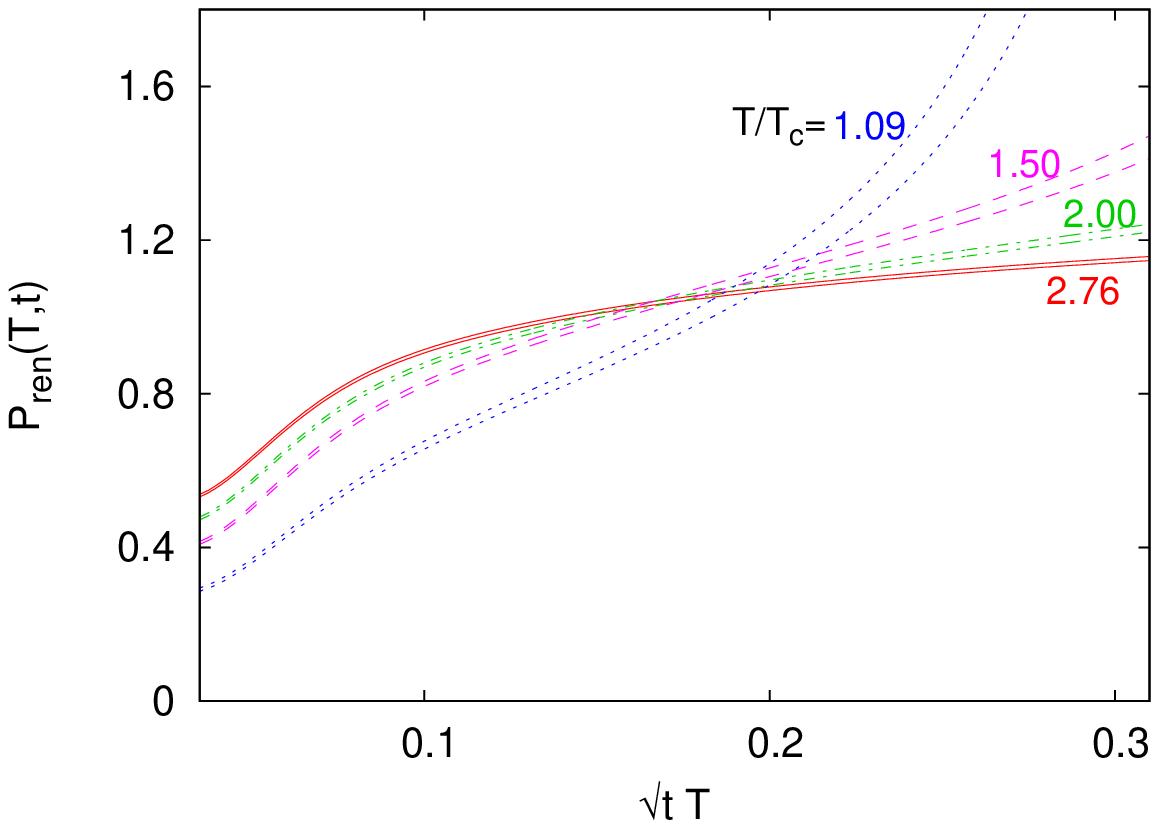}
\caption{(Left)$\prent$, \eq{expr} as function of flow time, at a
  temperature of 2 $T_c$ and different lattice spacings. Shown are the
  results using $\gms$ (solid lines, see below), $\gms$ extracted from
  $\ebf$ using \eq{condpert} (dashed lines, denoted
  $g^2_{\scriptscriptstyle{\rm MS, flow}}$) and $g^2$ in the flow
  scheme(dotted lines). The thickness of the lines shows the error.
  (Right) $\prent$ at a few temperatures on $N_t=8$ lattices. Here the
  two-loop $\overline{MS}$ coupling has been used. The two lines of the
  same style define the error band.}
\eef{polrenfac}

We first explore perturbative renormalization, using \eq{expr}.  The
coupling $\gms$ is evaluated using the two-loop formula with $\lms \,
= \, 1.20(2) \; \tc$ \cite{largen}, determined using plaquette values
and two-loop perturbation theory.  This value agrees, within error
bars, with the value $1.24 \pm 0.10$ quoted in Ref. \cite{tct0} and
values in the range 1.18-1.22 obtained in Ref. \cite{flow2}, as well as an
earlier measurement of $1.15 \pm 0.05$ \cite{gupta}. Note that the
starting point of the calculation of $\lms$ is a lattice observable,
and in the references cited above, two-loop perturbation theory was
used to extract $\lms$; so it is only consistent to calculate the
coupling using the two-loop formula.

In the left panel of \fig{polrenfac} we illustrate the perturbative
renormalization by showing $\prent$ at 2 $\tc$ at different lattice
spacings.  One clear lesson from this exercise is that when $b
\lesssim 1/N_t$, the multiplicative renormalization does not work.
This follows from our earlier observation that there remains an
$\mathcal{O}(a^2/t)$ piece which breaks scaling. That this should be
large seems reasonable when one remembers that at such values of $b$
one has $\sqrt{t}/a \lesssim 1$.  In this region of flow time there is
a rapid rise in the value of the renormalized Polyakov loop. A much
milder dependence on $t$ is observed at larger $b$. In this figure we
also make a comparison with calculations where the coupling is
calculated differently, in particular, calculations where $\gms$ is
obtained through \eq{condpert} (this is denoted by $g^2_{\scriptscriptstyle 
{\rm MS, flow}}$ in the figure), and also where the renormalization
factor is calculated with the flow coupling $g^2_{\rm flow}$ (\eq{flowcoupling}). 
In each
case the thickness of the line shows the error bar, combining the
statistical error in the data and the uncertainty in the coupling. The
calculation in the flow scheme is seen to result in a stronger $t$
dependence. This may indicate that the higher order corrections are 
larger in the flow scheme \footnote{An explicit calculation showed
  this to be the case for QED, where, for one fermion flavor, one gets
\[
R \left(g^2 \right) \ = \ \frac{\alpha}{\sqrt{8 \pi}} \, 
\left( 1 + 0.36 \frac{\alpha}{\pi} \ + \ \mathcal{O}(\alpha^2) \right)
\]
where $\alpha=e^2/4 \pi$ stands for
$\alpha_{\scriptscriptstyle{\overline{MS}}} (\mu=1/\sqrt{8 t})$.
The flow coupling in this case is 
\[
\alpha_{\rm flow} \ = \ \alpha \left( 1 + 1.16 \frac{\alpha}{\pi} 
\ + \ \mathcal{O}(\alpha^2)\right).
\]
}. The results are very close (for $\sqrt{t} > a$)
in the two calculations where $\gms$ is obtained from the two-loop
perturbation theory using $\lms$ and where it is defined through
\eq{condpert}.  The coupling $g^2_{\scriptscriptstyle {\rm MS, flow}}$, 
obtained using \eq{condpert} from data at
non-zero $a$, will have finite lattice spacing effects, and
will also differ from other two-loop evaluations of $\gms$ at
$\mathcal{O}(g^6)$. The agreement in \fig{polrenfac}
indicates that such effects are small at these couplings. As $T$ is lowered,
the agreement at fixed $\sqrt{t}T$ becomes worse, as one would expect
from the increase in coupling.

In the right panel of the same figure, we show the perturbatively
evaluated $\prent$ at different temperatures, for lattices with
spacing $a=1/8T$.  In this, and all the following figures where a
perturbative coupling is used, the two-loop $\overline{\scriptstyle
  MS}$ coupling, calculated using $\lms \, = \, 1.20(2) \; \tc$, has
been used.  The growth of the error band at lower temperatures is
because of the increase in scale dependence of the two-loop
coupling. The knee at $\sqrt{t} \sim a$ can be seen at the different
temperatures.  For $\sqrt{t} > a$, the dependence of $\prent$ on
flow time is mild at high temperatures, but less so at lower
temperatures.

To illustrate the temperature dependence of $\prent$, we show it in
\fig{renflowT} at both $c=0.15$ and $b=0.2$.  The remnant $t$
dependence of $\prent$ is clear by comparing the two panels of the
figure.  The corresponding susceptibility densities are also shown in
the figure.  The value of $T_c$, defined by the susceptibility peak,
is consistent between the computations using different flow times.
However, the value of $\chil$ depends on the choice of $t$ and the
scheme.

\bef
\includegraphics[scale=0.7]{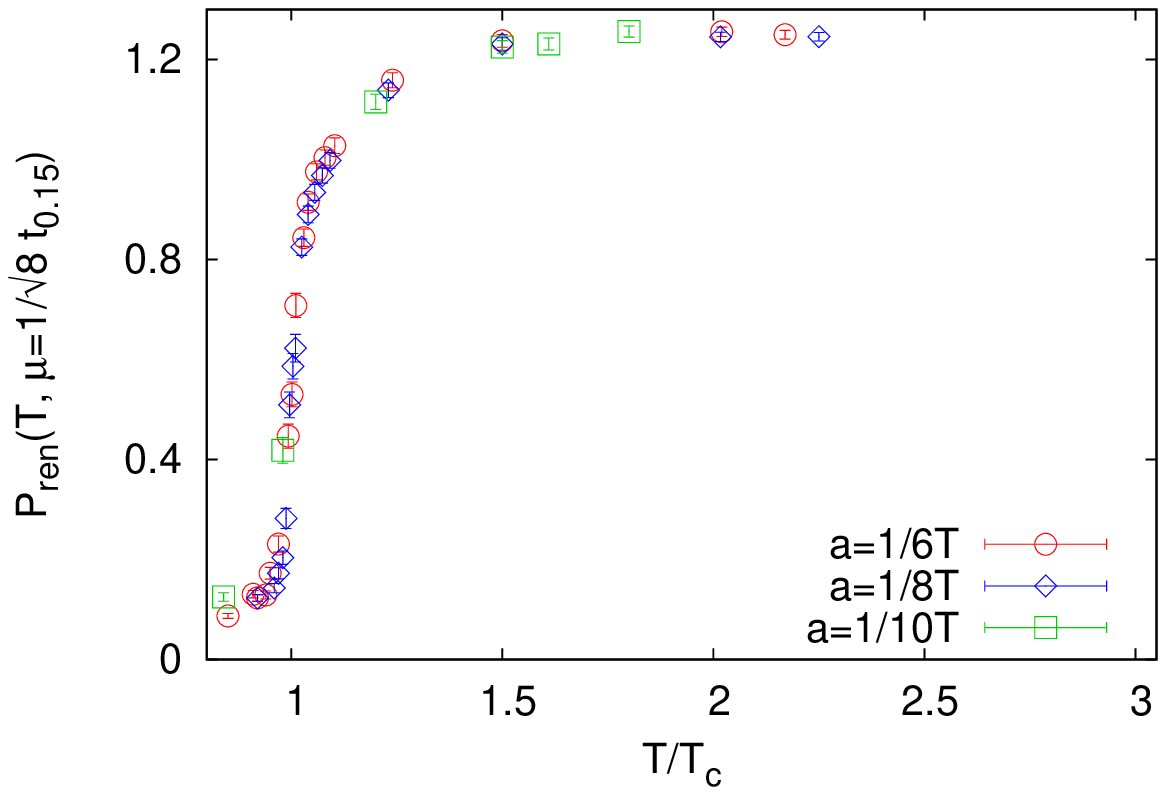}
\includegraphics[scale=0.7]{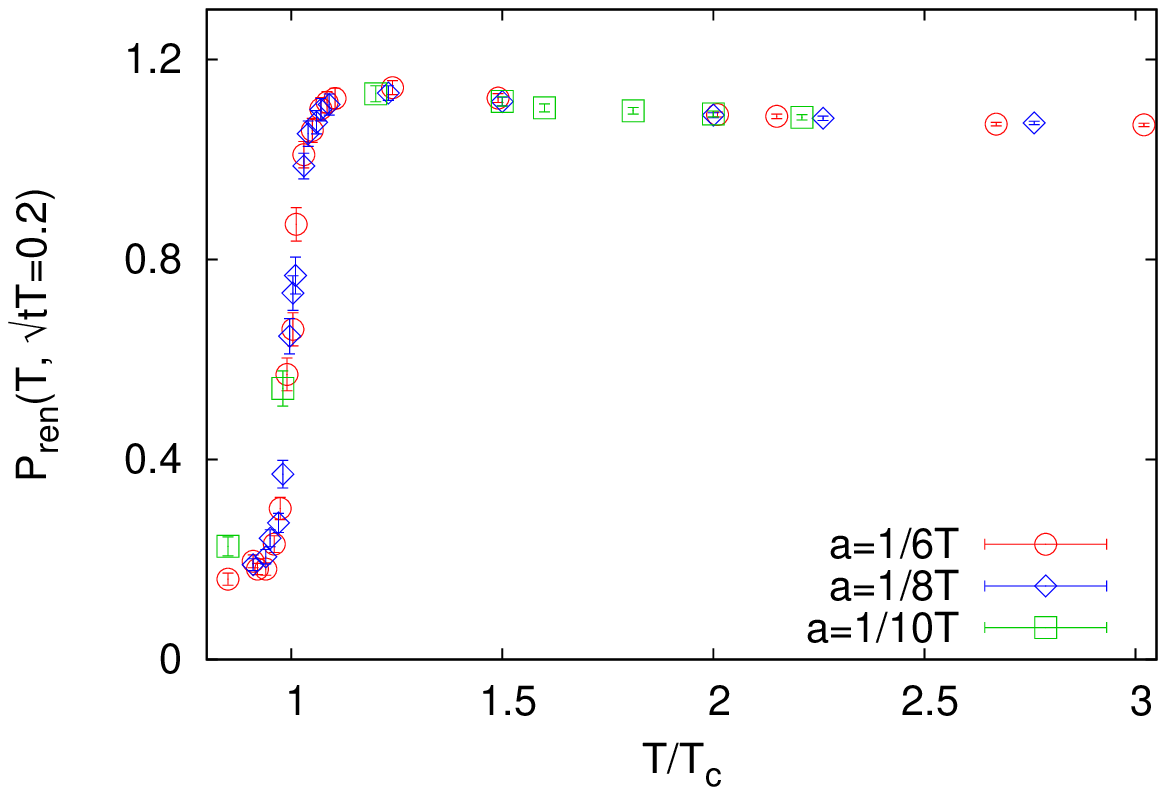}
\includegraphics[scale=0.7]{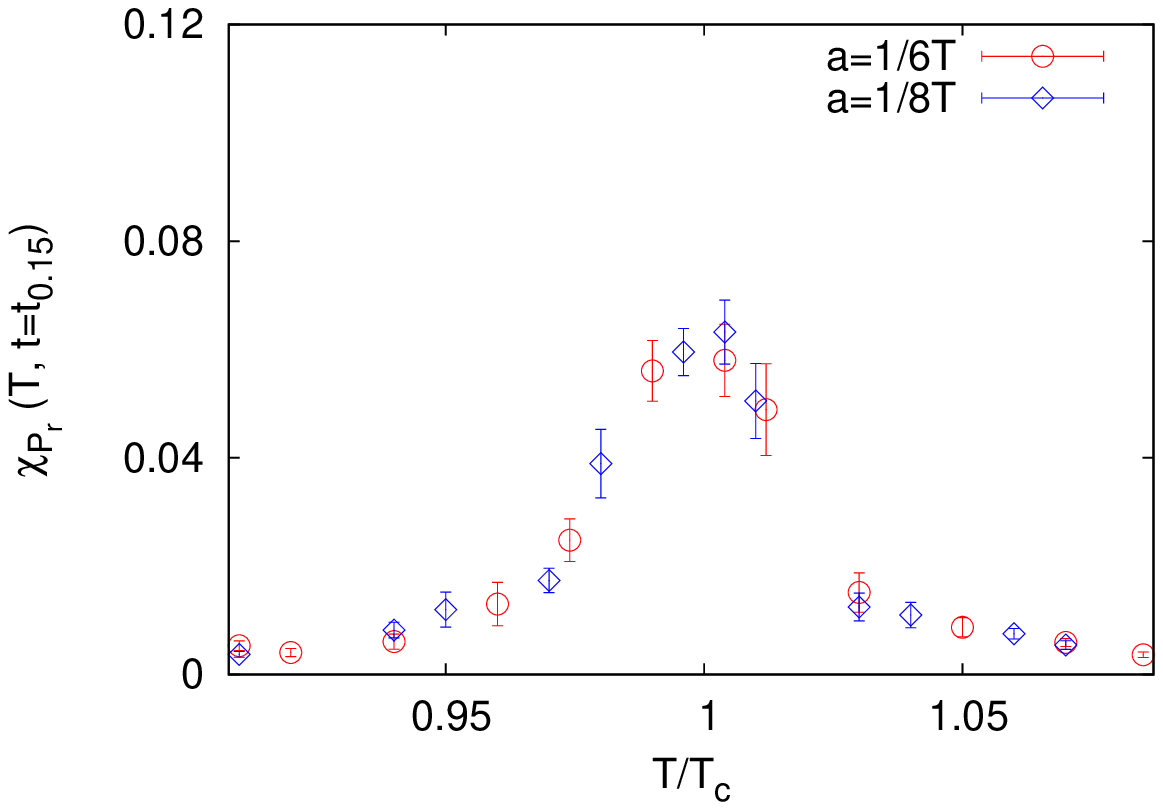}
\includegraphics[scale=0.7]{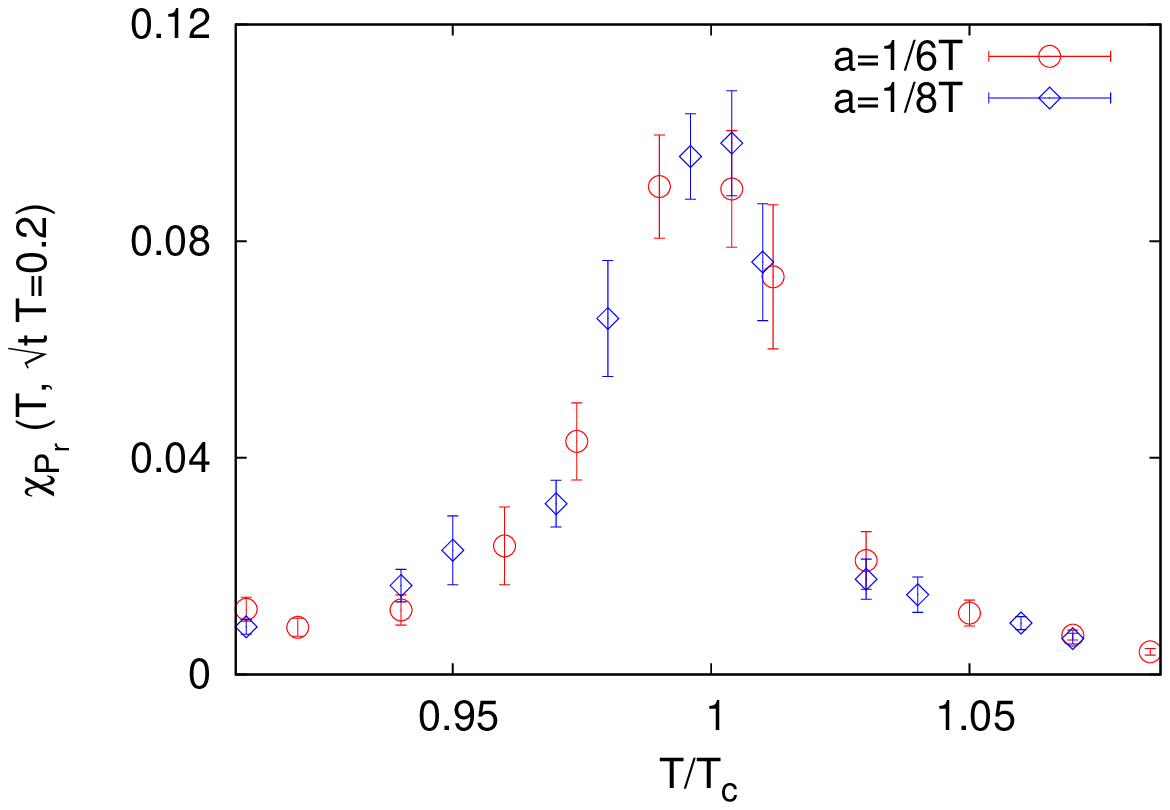}
\caption{The renormalized flowed Polyakov loop $\prent$ (top), at flow times
$t_{0.15}$ (left) and $t=(0.2/T)^2$ (right), and the corresponding 
susceptibilities (bottom).}
\eef{renflowT}

The substantial $t$ dependence in $\prent$ defined through \eq{expr},
in particular at lower temperatures, is not unexpected, as we are
using only a leading order renormalization factor in perturbation
theory. If one assumes that the $t$ dependence is due to
remnant $\mathcal{O}(g^4 \tT)$ effects, one can attempt a linear
extrapolation of $\prent$ to $t \to 0$. This is similar to the
strategy of Ref. \cite{flowqcd}. As \fig{polrenfac} reveals, such an
extrapolation is definitely not viable at smaller temperatures, where
the $t$ dependence is strong and complicated.  At higher temperatures,
a linear behavior does not set in at $\sqrt{t}/a \sim 1$, but a linear
extrapolation is feasible from a somewhat larger $t$.  As an
illustration of how the result of such a program will look, in the
left panel of \fig{polresult} we show the results of such an
extrapolation.  For definiteness, here we have chosen $\tT \in
(0.2,0.3)$ for all the fits. This choice of range was guided by the
discussion at the beginning of Section \ref{sec.flowpoly}, as well as
a preference for a fixed range for all lattices, and the requirement
that the result should not change, within errors, for a small change
of the range.

While the perturbative strategy is straightforward, as we have
discussed, it may work only at high temperatures $\gtrsim 2 \tc$.  By
going to higher $N_t$ it may be possible to make the extrapolation
more stable at lower temperatures; however, it will be difficult to
push it down to $\tc$ with realistic lattices.  A more viable,
nonperturbative strategy to calculate $\pren$ at temperatures close to
$\tc$ is to use the fact that the temperature dependence of the
renormalization factor is simple, \eq{prent}, and therefore, the
renormalization factor at one temperature can be simply obtained from
the renormalization factor at a different temperature modulo remnant
linear $\sqrt{t} T$ corrections, which we expect to be small if we
remain within our window $\sqrt{t} T \in (0.2, 0.3)$. In order to
extract $R \left( g^2(t) \right)$, we take a baseline value of the
Polyakov loop at a given temperature. In what follows, we take the
value $P_{\rm ren}(3 T_c)$ = 1.0169(1) \cite{renpol1} as the baseline.
This determines $R \left( g^2(\sqrt{t}=1/10 T_c) \right)$, which can
then be used to calculate $P_{\rm ren}$ to all temperatures up to 2
$\tc$. This process is then iterated to calculate $P_{\rm ren}$ at
lower temperatures.  This strategy is similar in spirit to that
followed in Ref.~\cite{renpol1}; however, the use of flow makes the
calculation simpler, as we do not need to match lattices at different
lattice spacings to the same temperature. The renormalized Polyakov loop
extracted this way is shown in the right panel of \fig{polresult}.

\bef
\includegraphics[scale=0.7]{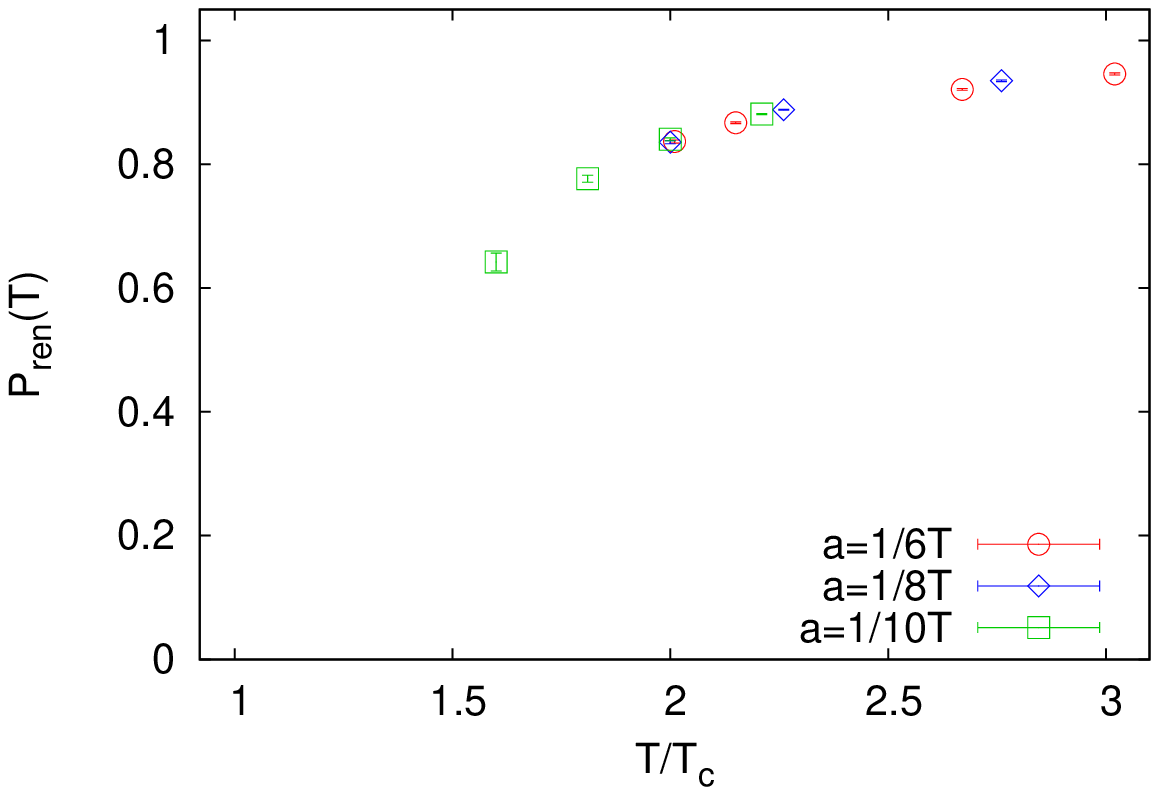}
\includegraphics[scale=0.7]{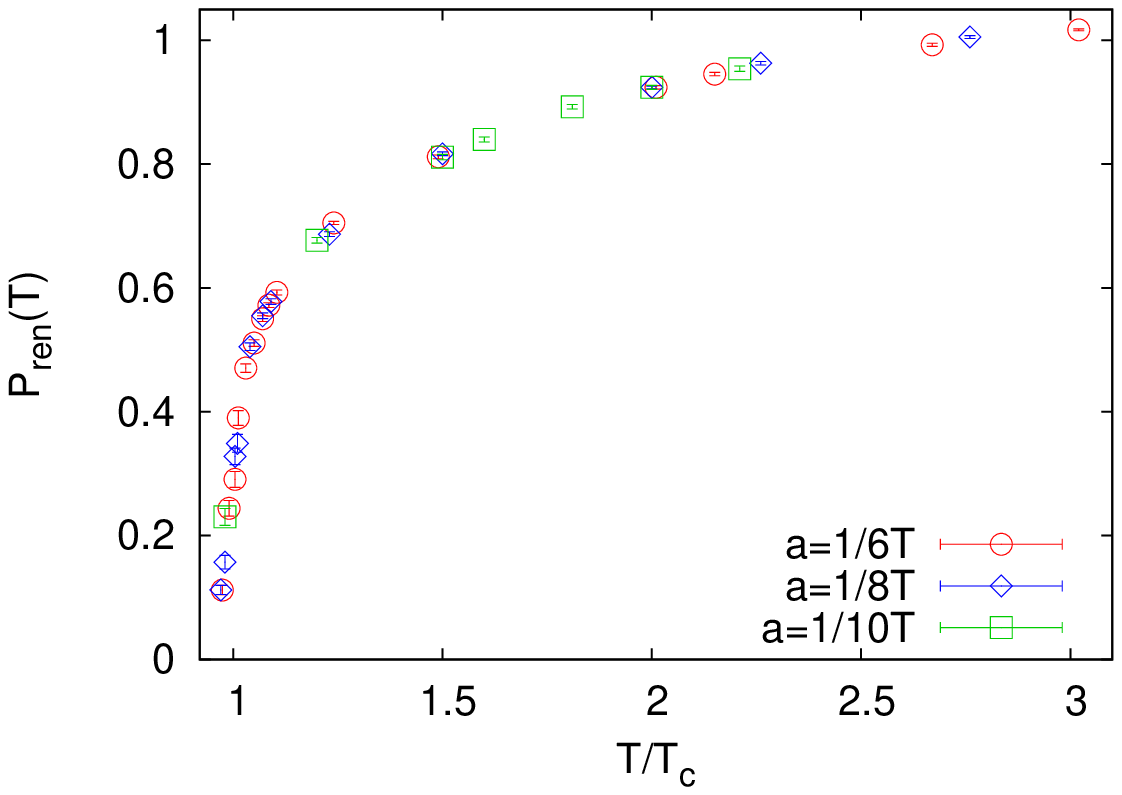}
\caption{Renormalized Polyakov loop $\pren$. (Left) Result of $t \to 0$
extrapolation of the leading order renormalized Polyakov loop, $\prent$
(\eq{expr}). (Right) Renormalized Polyakov loop where the renormalization
factor $R(g^2(t))$ is obtained from a nonperturbative matching; see text.}
\eef{polresult}

\section{Electric and magnetic condensate}
\label{sec.cond}
The nonperturbative nature of the QCD vacuum is characterized by various 
condensates, which melt across the deconfinement transition. In the pure glue 
theory at zero temperature, $\ebf$, \eq{cond}, is the only dimension four, 
scalar operator one can form. At finite temperatures Lorentz symmetry is 
broken, and two separate rotationally invariant, positive parity operators can 
be constructed out of $\ebf$:
\beq
E = \tr G_{0i} G_{0i}, \qquad M = \frac{1}{2} \tr G_{ij} G_{ij}
\eeq{ftcondop}
which are related to the electric and magnetic gluon condensates, respectively.
$O(4)$ symmetry at zero temperature implies $E$ and $M$ are
not independent operators, and in the rest frame, 
$\langle E \rangle = \langle M \rangle = \frac{1}{2} 
\langle \ebf \rangle$. 
 
The flow behaviors of $E$ and $M$ turn out to be quite interesting. 
In \fig{flowcond} we show the dimensionless flowed quantities 
$t^2 \Eav$ and $t^2 \Mav$ immediately below and above $T_c$. 
In the same figures we also show the flow behavior of the 
same operator at $T=0$, studied on $N_s^4$ lattices at the same $a$.
Below $T_c$ the flow time behavior of the operators is
identical, indicating that even at $0.92 \tc$, O(4) symmetry is 
approximately satisfied in the pure glue theory. 
In contrast, just above $\tc$, the flow behavior of $E$ and $M$ turn
out to be very different from each other. While at small $t$, the flow
behavior is influenced by the lattice cutoff and is similar to that
seen for $\ebf$, at longer flowtime the growth of $\Eav$ with flowtime
flattens out while $\Mav$ grows rapidly. Note that this behavior sets
in in a very narrow region around $\tc$. While the breaking of the
O(4) symmetry can already be seen at tree level, the interacting
theory shows a much stronger effect (see Appendix \ref{cond.lo} and
\fig{lo.cond}). This dynamical realization of
the O(4) symmetry for all $T < T_c$, and its abrupt breaking just
above, is consistent with the observation that both the energy density
and the interaction measure vanish below $T_c$ and are finite just
above. A similar realization of O(4) symmetry at finite temperature
below $T_c$ was also observed in the screening of glueball-like
operators \cite{glueball}.

In \fig{flowconddiff} we show the difference $t^2 \langle E(T,t) -
M(T,t) \rangle$ .  The panel on the left shows the flow-time behavior
for different $T$.  For $\tT \gtrsim 1/N_t$ this quantity is very
sensitive to the deconfinement transition. For $T < T_c$ the quantity
remains small.  However, for $T > T_c$ significantly larger values are
observed. Note that the $1/t^2$ singularity is cancelled between the
electric and the magnetic operator expectation values. This allows us to
study the difference $\langle E(T, t) - M(T, t) \rangle /T^4$.

Figure \ref{fig.flowconddiff} shows $\langle E(T, t) - M(T, t) \rangle /T^4$ 
as a function of $T/T_c$. At $t = 0$ this is a multiple of the entropy
density, which is known to change abruptly across the pure gauge
transition. At larger $t$ this jump is even more pronounced. What we
would like to emphasize here is that for the flowed operator, this
sharp jump arises from the flow behavior of $\Eav$ and $\Mav$. 
The flowed $\langle E - M \rangle /T^4$ can be used as an additional 
marker for the deconfinement transition.
 
\bef
\centerline{\includegraphics[scale=0.5]{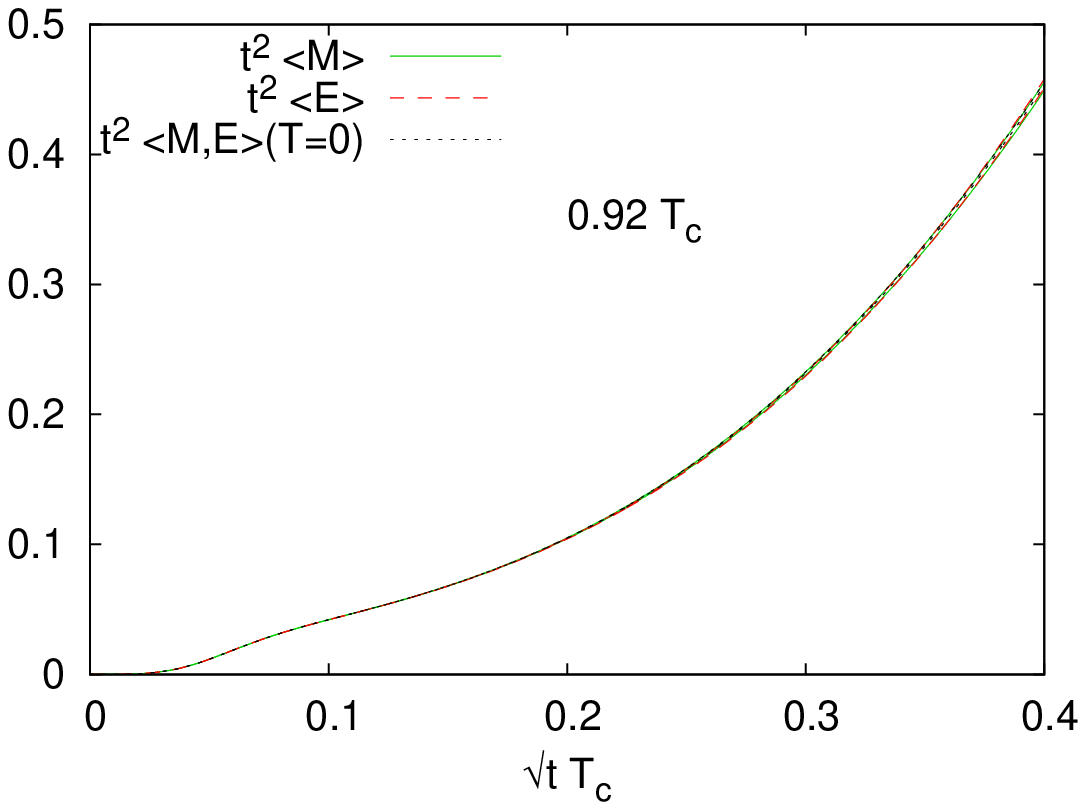}
\includegraphics[scale=0.5]{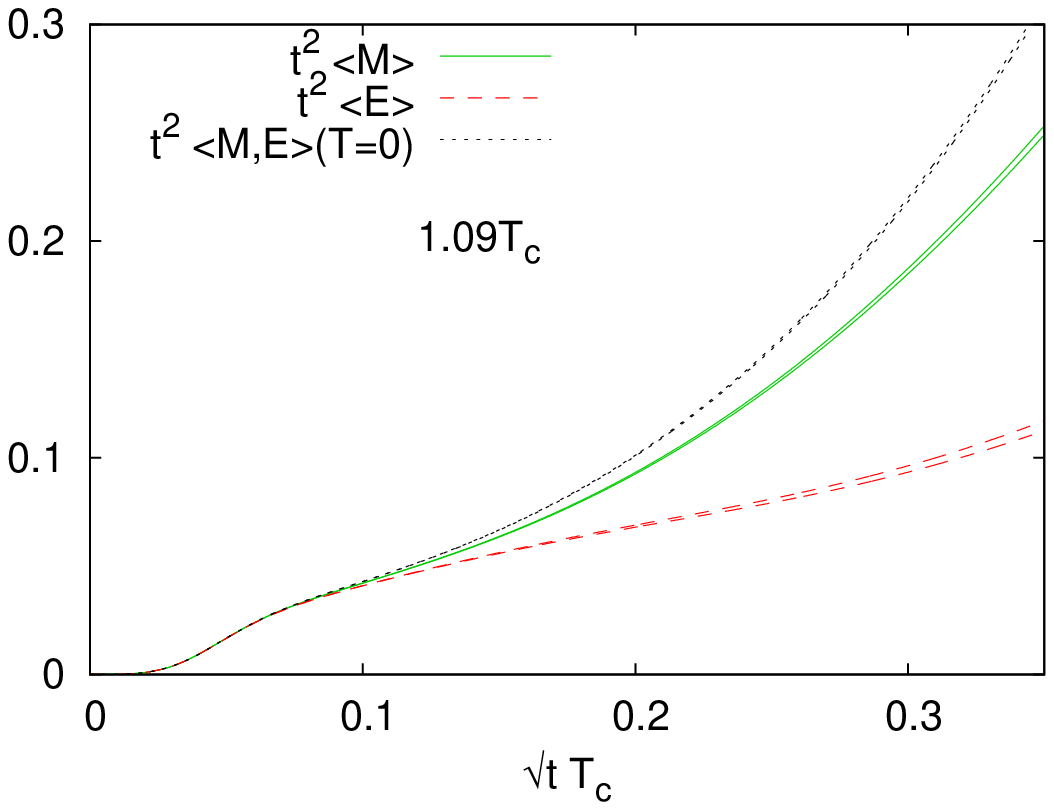}
\includegraphics[scale=0.5]{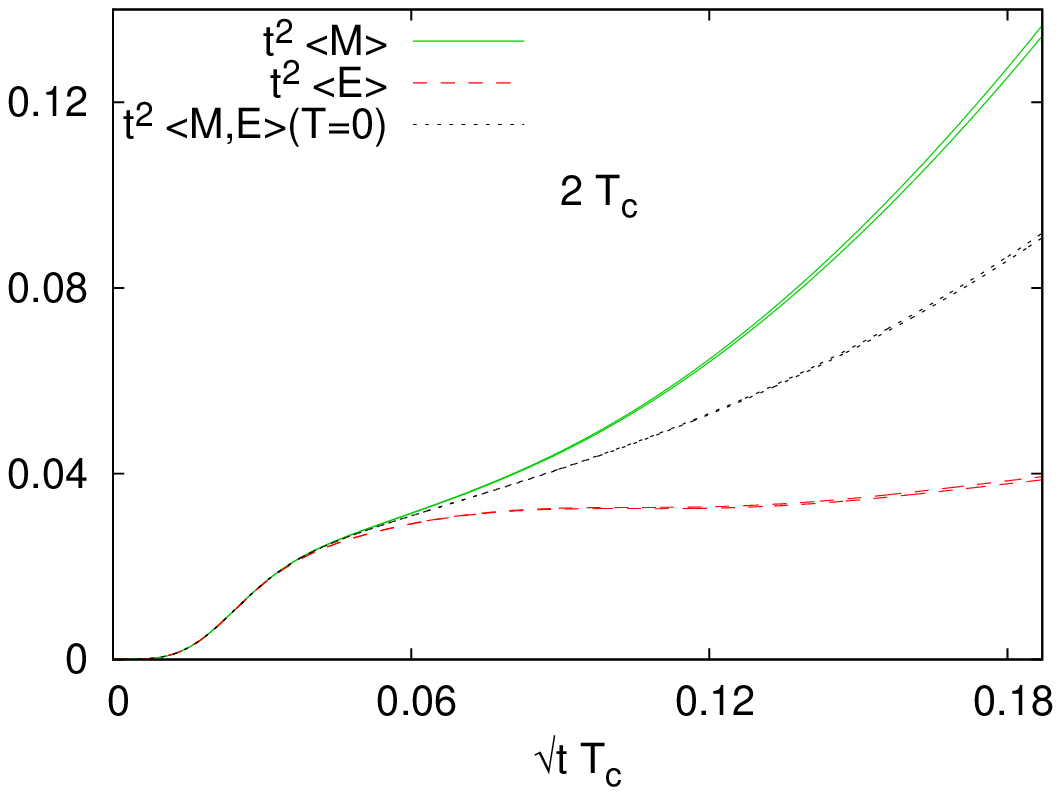}}
\caption{The electric and magnetic condensate operators $t^2 \langle
  M, E \rangle$ plotted against flow time, at temperatures of 0.92
  $T_c$ (left), 1.09 $T_c$ (middle), and 2 $T_c$ (right), on lattices
  with $a=1/8T$. We show the $1 \sigma$ band of the quantities.  Also
  plotted are the zero temperature values of the same operators. In
  the left panel ($T < T_c$) the three bands almost completely
  overlap.}
\eef{flowcond}

\bef
\includegraphics[scale=0.7]{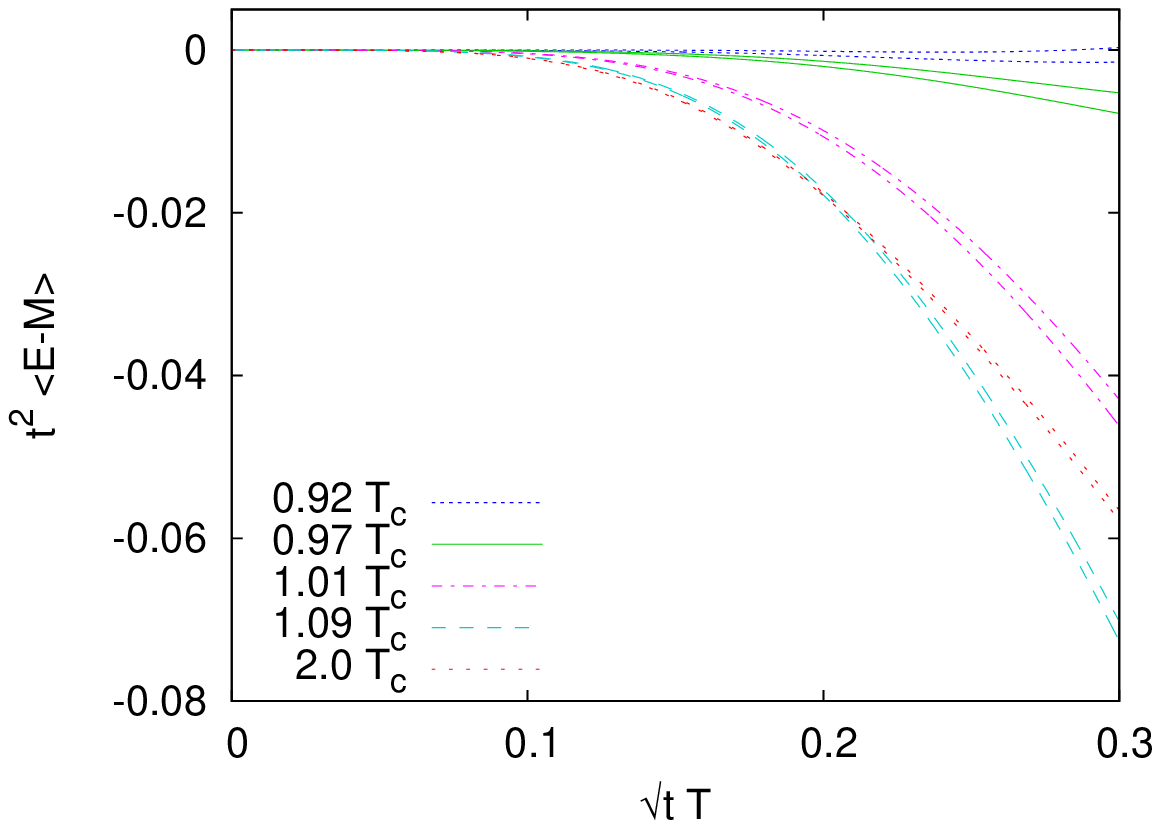}
\includegraphics[scale=0.7]{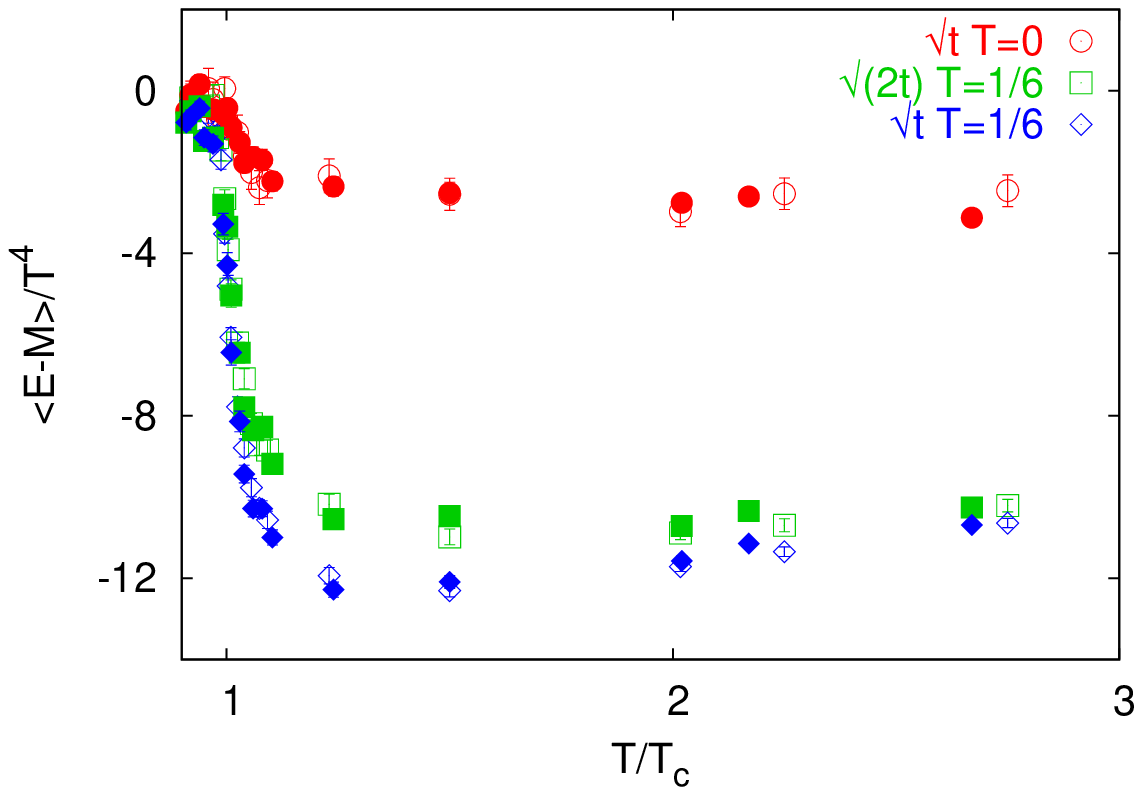}
\caption{(Left) Flow time dependence of the operator $t^2 \langle E -
  M \rangle$ at different temperatures, for lattices with
  $N_t=8$. The two lines of the same style define the 1 $\sigma$ band.
(Right) Temperature dependence of the condensate difference
  $ \EMav /T^4$, without the flow and for two
  different flow times. Shown are the results for two different sets:
  the empty symbols correspond to $N_t=8$ lattices while the filled
  symbols correspond to $N_t=6$ lattices.}
\eef{flowconddiff}

\subsection*{Connection to electric and magnetic gluon condensates}

The vacuum gluon condensate is defined through the expression
\footnote{Another common convention uses the multiplicative factor
  $\beta(g) / g^3$. Our convention is 8/11 times smaller.} \\
\beq
\Gsq = \left\langle \frac{ 8 \beta(g)}{11 g^3} \ \tr G_{\mu \nu} G_{\mu \nu} (T=0) 
\right\rangle_{\rm subt}\ = \left\langle \frac{1}{2 \pi^2} \, 
(1+\mathcal{O}(g^2)) \,  \tr G_{\mu \nu} G_{\mu \nu} (T=0)
\right\rangle_{\rm subt} ,
\eeq{gcond}
where $\beta(g) \, = \, \mu \frac{\textstyle \partial g}{\textstyle \partial \mu} 
\, = -b_0 g^3 - b_1 g^5 - ..., \ \ b_0 \ = 
\ \frac{\textstyle 11}{\textstyle 16 \pi^2}.$ We have earlier used
the quartic divergence of $\langle \ebf \rangle$ to define the flow
scale.  Here, the subscript on the vacuum expectation value (VEV) 
indicates that the hard mode
contribution has been subtracted off.  The resulting expectation value
is finite and quantifies an important nonperturbative property of the
vacuum \cite{svz}.

Much effort has gone into the extraction of this property of QCD
from either experiment or lattice calculations, but the extraction is still not
stable.  From analysis of the decay of the $\tau$ lepton, a value of
0.02-0.01 ${\rm GeV}^4$ has been quoted for the gluon condensate
\cite{donoghue} using a subtraction point $\sim (2 \ \GeV)^4$, while a
recent determination quoted $0.009 \pm 0.007 \ \GeV^4$ \cite{ioffe}. For 
SU(3) gauge theory a recent lattice determination
quotes $\Gsq = 24.2 \pm 8.0 \ \Lambda_{\overline{MS}}^4$ \cite{bali},
where $\Gsq$ has been defined after subtraction of the perturbative
part.
 
A finite temperature gluon condensate was defined analogously
\cite{leutwyler} by merely replacing the VEV in \eq{gcond} by the
thermal expectation value: 
\beq \GsqT = \left\langle \frac{16
  \beta(g)}{11 g^3} \ \ebf \right\rangle_{T, \rm subt} .
\eeq{ftcond} 
The complication of the hard mode subtraction can then be avoided by
studying $\Gbar = \GsqT - \Gsq$. This difference is obtained simply
from the difference of the expectation value of $\ebf$ on a thermal
and a zero temperature lattice at the same $a$.  $\Gbar$ is
proportional to the trace of the energy-momentum tensor, and has been
calculated from the plaquette operators for SU(3) \cite{su3thermo}. In
addition, we can also define the electric and magnetic gluon
condensates, $\GesqT$ and $\GmsqT$, analogously by replacing $\ebf$ by
$E$ and $M$ respectively in \eq{ftcond} \cite{su3thermo}.  Finite
temperature sum rule calculations use all these condensates
\cite{lee}.

The connection between the flowed condensate operators
and the electric and magnetic gluon condensates can be extracted from
Ref.~\cite{suzuki}: \\
\beq
\left. \begin{aligned}
\Gbar \\
\Ebar \\
\Mbar
\end{aligned} \right\} \ = \lim_{t \to 0} \ R(t) 
\cdot \left\{ \begin{aligned} 
\langle \ebf(T, t) \rangle - \langle \ebf(T=0, t) \rangle \\
\langle E(T, t) \rangle - \langle E(T=0, t) \rangle \\
\langle M(T, t) \rangle - \langle M(T=0, t) \rangle \\
\end{aligned} \right.
\eeq{condflow}
where the renormalization constant $R(t)$ at leading order is \\
\beq
R(t) = \frac{1}{\pi^2} \ \left(1 - 2 \, b_0 \, \bar{s}_2 \, 
g^2(\mu=1/\sqrt{8} t) \ + \ \mathcal{O}(g^4) \right)
\eeq{condrenorm}
with $\bar{s}_2$ = 0.055785 \cite{suzuki}.

\bef
\includegraphics[scale=0.7]{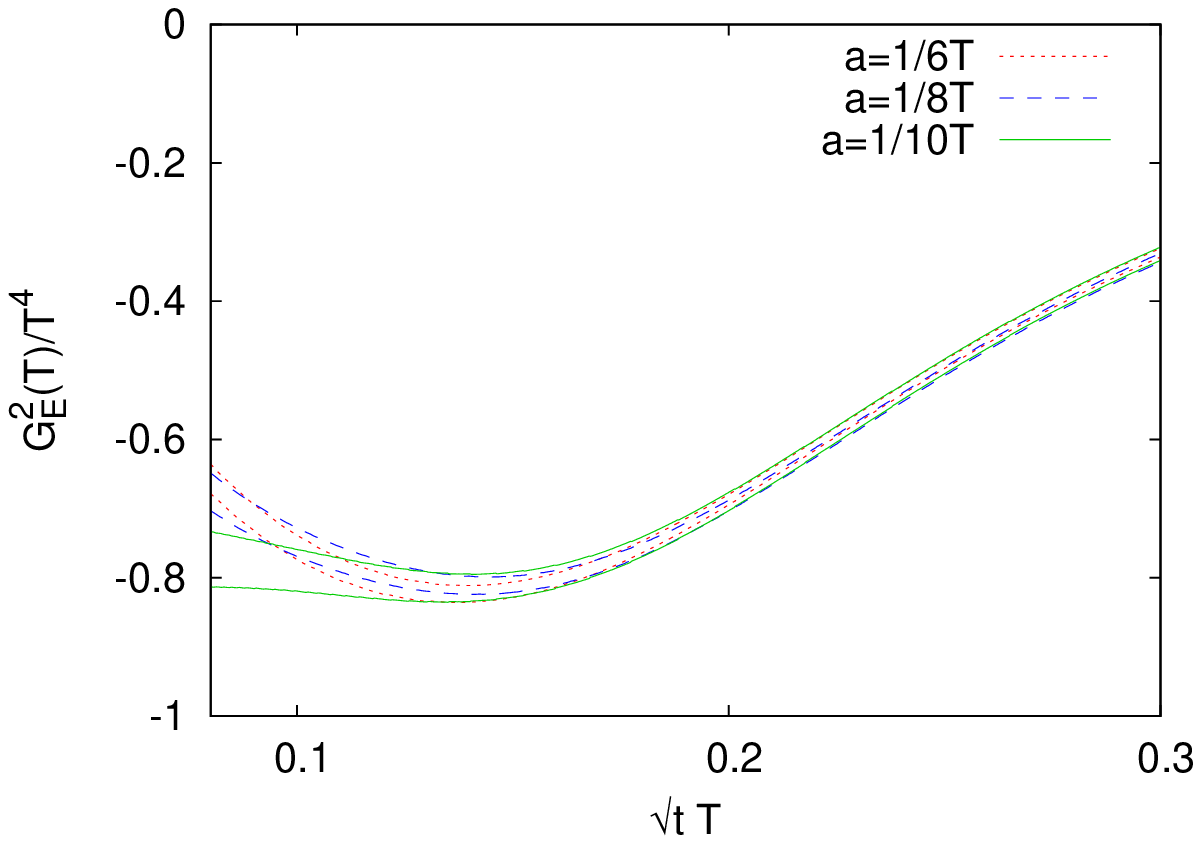}
\includegraphics[scale=0.7]{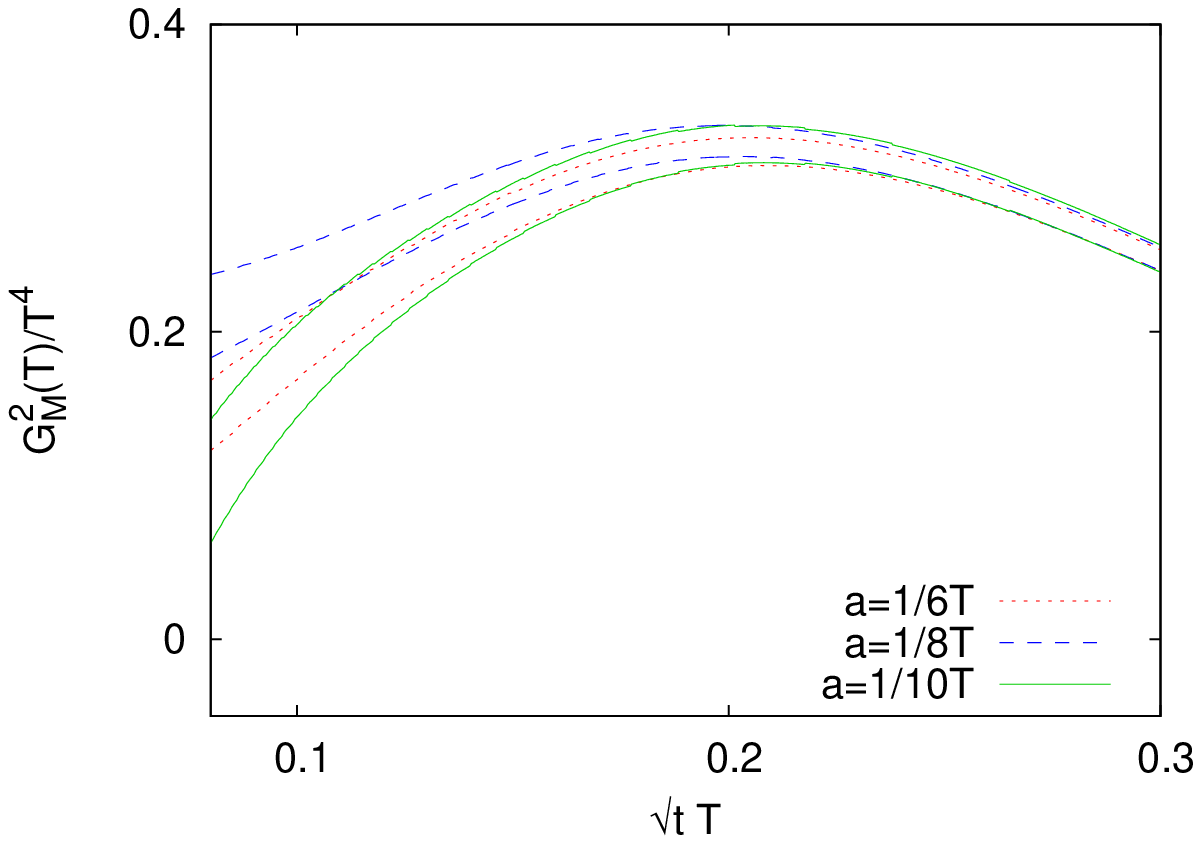}
\caption{The renormalized condensates $\Ebar$ (left) and $\Mbar$ (right)
at $2 T_c$, plotted as functions of flow time, at three different 
lattice spacings. The two lines of the same style define the error band.}
\eef{flowrenormcond}

In \fig{flowrenormcond} we show the renormalized condensates $\Ebar$
and $\Mbar$ as a function of flow time. Their sensitivity to the flow
time means that the values of the renormalized condensates depend on
the scale at which they are extracted.  As we discussed before, the
extraction of thermal physics from flowed configurations will require
the flow time $\sqrt{t} T$ to be in a small window: $0.16 \lesssim
\sqrt{t} T \ll 1/\sqrt{8} \sim 0.35$. While the lattice spacing
dependence is small within this window, neither a clear linear behavior,
nor a prominent plateau can be seen for us to reliably extract the 
condensates using \eq{condflow} \footnote{This may be observable dependent.
For related observables a plateau was reported in \cite{flowqcd} even 
close to $\tc$.}. There is a hint of a plateau near the lower end of the 
window; in order to get a qualitative idea of the temperature 
dependence of the condensates, in \fig{allcond} we show the renormalized 
condensates from this region, at the flow time $\sqrt{t} T = 1/6$.
As the discussion here suggests, this is to get only a 
qualitative idea of the dependence of the condensates on temperature. 

\bef
\includegraphics[scale=0.7]{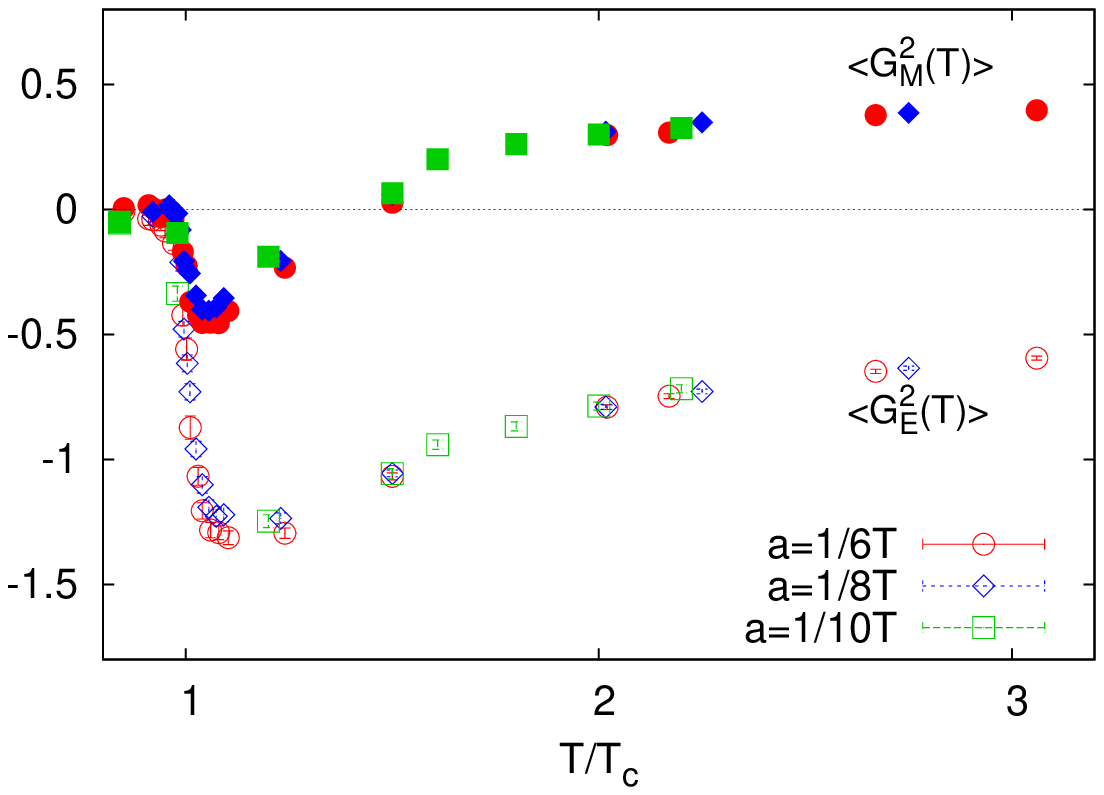}
\includegraphics[scale=0.7]{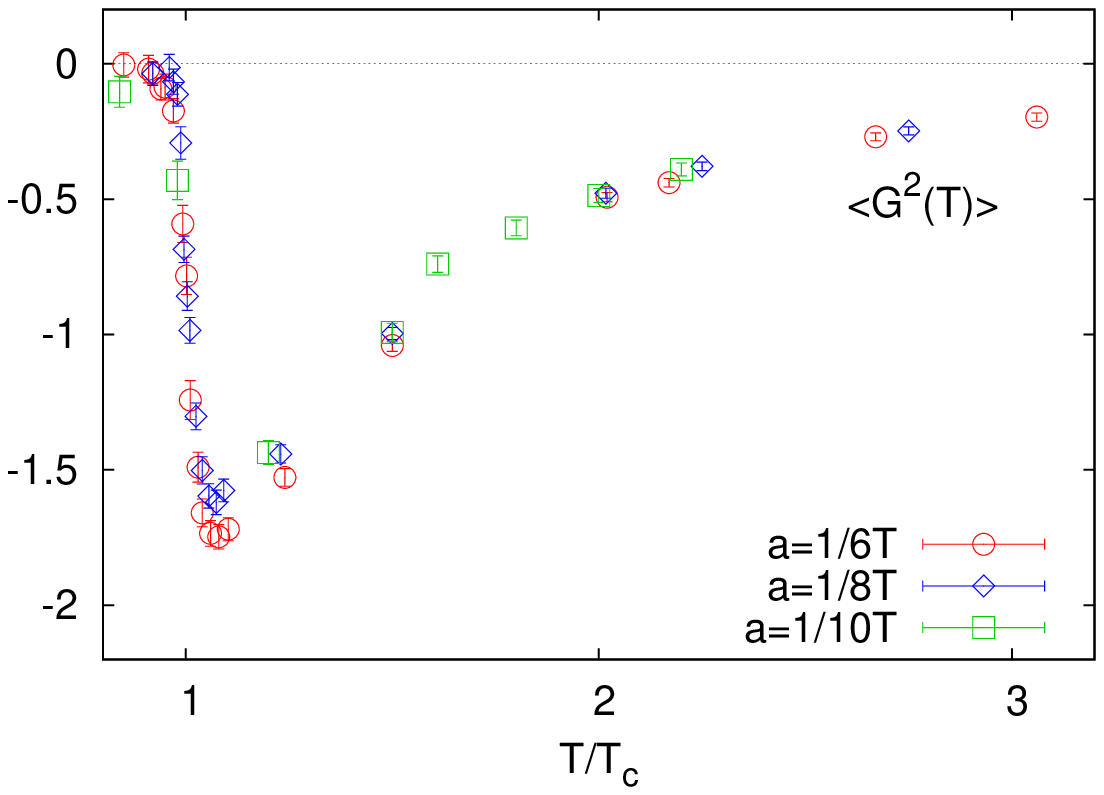}
\caption{(Left) The renormalized condensates $\Ebar$ (open symbols) 
and $\Mbar$ (filled symbols) 
plotted as a function of temperature, at flow time $\sqrt{t}T=1/6$.
(Right) The corresponding plot for $\Gbar$.}
\eef{allcond}

The figure shows some interesting features. Both $\Ebar$ and $\Mbar$
grow rapidly in magnitude just above the deconfinement transition,
with $\Ebar$ increasing more rapidly. The maxima of their magnitudes
occur at approximately the same temperature, just above $\tc$. This
leads to a sharp dip in $\Gbar$. After this temperature, the magnitude
of both $\Ebar$ and $\Mbar$ decrease, while their difference is more
stable. This causes the near-flat temperature dependence of $\EMav$
seen in \fig{flowconddiff}.  Eventually $\Mbar$ changes sign (see also
\fig{flowcond}). In the leading order, $\Ebar$ = - $\Mbar$
(\eq{lo.condsubt}). An indication of the approach to such behavior is
already seen in our highest temperatures.  This leads to a very small
value of $\Gbar$ at these temperatures, and therefore a very sharp
drop after the peak just above $\tc$. Qualitatively, the behaviors are
similar to that seen in Ref. \cite{su3thermo} for analogous operators 
calculated from the plaquette, and
renormalized using a nonperturbative $\beta$ function.

\section{Discussion and conclusion}
\label{sec.summary}
In this paper we have discussed the use of Wilson flow to study the
deconfinement transition in SU(3) gauge theory. In particular, we have 
emphasized the construction of operators for studying the onset of 
the deconfinement transition. While the explicit computations done here are
for the pure gauge theory, we expect the qualitative features to be true 
also for the theory with fermions. Therefore Wilson flow has the promise of 
being a powerful diagnostic tool for the deconfinement transition in QCD.

In Sec. \ref{sec.poly} we have discussed flowed Polyakov loops, i.e.,
Polyakov loops constructed from links flowed to a fixed physical
distance.  Since the flow preserves Z(3) symmetry, and flowed Polyakov
loops do not require renormalization, they give renormalized order
parameters for the deconfinement transition. We have investigated both
the cases of the flowed distance being fixed in terms of temperature
or in terms of a temperature-independent physical length scale. Their
behaviors are discussed in Sec. \ref{sec.flowpoly} and shown in \fig{flowT}.

From the flowed Polyakov loops, renormalized (thin) Polyakov loops can
be obtained. The renormalization using leading order perturbation
theory, illustrated in \fig{renflowT}, has remnant dependence on flow
time, in particular close to the transition temperature.  This is not
entirely unexpected, as for the flow times we can use (which is
restricted below by the lattice spacing $a$ of our lattices), the
coupling is not small. We therefore turned to a nonperturbative
renormalization. The results for the renormalized 
Polyakov loop are shown in \fig{polresult}. The use of the flow allowed us 
to conveniently do the nonperturbative matching involved in the 
renormalization. In principle, one can use the flow to do a purely 
perturbative renormalization also; the result of such a computation, 
with the leading order renormalization constant, is also shown. Perhaps
not surprisingly, the perturbative method does not work very well 
especially as one comes down in temperature. Both higher order calculations
and much finer lattices may be required for the perturbative approach to work 
near $\tc$.  

Next we turn to a discussion of condensates of gluonic operators. At
finite temperatures, two rotationally invariant, positive parity
operators of dimension four can be constructed, corresponding to
electric and magnetic gluon operators, \eq{ftcondop}. We find that
their flow behaviors are very sensitive to the deconfinement
transition: just around $\tc$ the flow behavior of the electric
condensate changes drastically. As a result, the difference between
the flowed operators acts as a marker of the transition. This is
illustrated in \fig{flowconddiff}.

An extraction of the renormalized gluon condensate from these results
is difficult, because there is a dependence on flow time which is not
cured by the leading order perturbative renormalization constant. Again
it is not a surprise that leading order perturbation theory does not
work at the lattices used by us. There is a hint of a mild plateau near
the lower end of the window in flow time where thermal physics can 
be extracted. We use the results from this region to illustrate qualitative
features of the  thermal behavior of the renormalized gluon condensates 
in \fig{allcond}. The figure 
shows interesting thermal behavior of the electric and magnetic condensates, 
in particular just above the deconfinement transition. 

{\bf Acknowledgements}: This work was carried out under the umbrella
of ILGTI.  The computations reported here were performed on the gaggle
cluster of the Department of Theoretical Physics, TIFR. We would like
to thank Ajay Salve and Kapil Ghadiali for technical support, and
Siddhartha Bhattacharyya, Anirban Lahiri and Shiraz Minwalla for discussions.

\appendix
\section{Calculational details}
\label{sec.detail}
For the results described in this paper, we calculated thermal and vacuum 
expectation values by generating zero and 
finite temperature lattices using a heatbath-overrelaxation algorithm.
For the finte temperature runs, $N_s^3 \times N_t$ lattices were generated
with $N_t \ll N_s$. Thermal averaging requires the gauge fields to be 
periodic in the Euclidean time direction. We also imposed periodic boundary 
conditions in the spatial directions. At each value of the gauge coupling, 
we also performed a zero temperature run with $N_s^4$ lattices.

One Cabibo-Marinari pseudoheatbath step was followed by three
overrelaxation steps; we call this combination a
sweep. Autocorrelations get enhanced by Wilson flow
\cite{milc,staggered}. To avoid autocorrelations, configurations were
separated by a large number of sweeps: 500 sweeps for the finite
temperature lattices and 200-500 sweeps for the zero temperature
lattices. The Wilson action was used for the gauge fields. A complete list
of the generated $T>0$ lattices is given in Table \ref{tbl.lattice}.

\begin{table}[tbh]
\begin{center} \begin{tabular}{|lll|lll|lll|} \hline
\multicolumn{3}{|c|}{$24^3 \times 6$} & \multicolumn{3}{|c|}{$32^3 \times 8$} &
\multicolumn{3}{|c|}{$32^3 \times 10$} \\
$\beta$ & \# conf & $T/T_c$ & $\beta$ & \# conf & $T/T_c$ & $\beta$ & \# conf  & $T/T_c$ \\ 
\hline
5.80 & 107 & 0.85 & 6.00 & 99 & 0.91 & 6.10 & 100 & 0.85 \\
5.84 & 105 & 0.91 & 6.02 & 100 & 0.94 & 6.20 & 100 & 0.98 \\
5.85 & 101 & 0.92 & 6.03 & 100 & 0.95 & 6.34 & 100 & 1.20 \\
5.86 & 101 & 0.94 & 6.04 & 100 & 0.97 & 6.50 & 100 & 1.50 \\
5.87 & 101 & 0.96 & 6.05 & 100 & 0.98 & 6.55 & 100 & 1.60 \\
5.88 & 101 & 0.97 & 6.06 & 199 & 0.996 & 6.65 & 100 & 1.81 \\
5.89 & 105 & 0.99 & 6.065 & 199 & 1.004 & 6.73 & 100 & 2.00 \\
5.895 & 105 & 1.004 & 6.07 & 199 & 1.01 & 6.80 & 100 & 2.21 \\
5.90 & 105 & 1.012 & 6.08 & 100 & 1.03 &       &   & \\
5.91 & 101 & 1.03 & 6.09 & 100 & 1.04 &      &    & \\
5.92 & 101 & 1.05 & 6.10 & 100 & 1.06 &      &    & \\
5.93 & 101 & 1.07 & 6.11 & 100 & 1.07 &      &    & \\
5.94 & 101 & 1.085 & 6.12 & 99 & 1.09 &      &    &  \\
5.95 & 105 & 1.10 & 6.20 & 99 & 1.23 &      &     & \\
6.02 & 105 & 1.24 & 6.34 & 100 & 1.50 &      &    & \\
6.14 & 105 & 1.49 & 6.55 & 103 & 2.00 &     &    &  \\
6.34 & 123 & 2.00 & 6.65 & 100 & 2.26 &      &   &  \\
6.40 & 105 & 2.15 & 6.80 & 100 & 2.76 &     &    & \\
6.55 & 105 & 2.67 &      &     &      &  &   &  \\
6.65 & 105 & 3.02 &      &     &      &  &   &   \\
\hline
\end{tabular}\end{center}
\caption{List of finite temperature lattices generated for our calculations.
Two configurations were separated by 500 x (1 HB + 3 OR) sweeps. For each
finite temperature set on lattices $N_s^3 \times N_t$, a set of $N_s^4$ 
lattices at the same $\beta$ was generated for the vacuum ensemble.}
\label{tbl.lattice} \end{table}

For generating the Wilson flow, the fourth order Runge Kutta was used 
\cite{main}. See Ref.~\cite{staggered} for an analysis of the convergence 
of this
scheme. We used $dt=0.01$ for our runs. For setting the temperature scales, 
the known results for $\beta_c$ at different $N_t$ were used, and $t_{0.12}$ 
was used for the relative scale at other $\beta$.

\section{Leading-order expressions}
\label{sec.lo}
In the body of the paper we have referred several times to the behavior
in leading order of perturbation theory. Here we write down the 
relevant expressions to this order for the Polyakov loop and the gluon 
condensates. 

\subsection{Polyakov loop}
\label{pol.lo}
In leading order, the flowed Polyakov loop is given by
\beqa
\log P(T, t) & = & - \frac{g^2}{4 N_c} \, \int_0^\beta d\tau_1 \, 
\int_0^\beta d\tau_2 \, \langle B^a(\tau_1, \vec{r}) \, 
B^a(\tau_2, \vec{r}) \rangle \nonumber  \\
& = & - \frac{g^2}{4 \pi^2} \cf \frac{\sqrt{\pi}}{\sqrt{8 t} T} 
+ \frac{g^2 \, \cf \, \me}{8 \pi T}  e^{z^2} \ \erfc(z)+ \mathcal{O}(g^4) 
\eeqa{pol}
where $\cf = \frac{\textstyle N_c^2 -1}{\textstyle 2 N_c}$,  
$z = \me \sqrt{2 t}$, with the electric mass $\me = g T$ for 
SU(3) gauge theory, and
$\erfc(z) = \frac{\textstyle 2}{\textstyle \sqrt{\pi}} 
\int_{\scriptstyle z}^{\scriptstyle \infty} e^{\scriptstyle -x^2} dx$. Therefore 
\beq
R \left(g^2(t) \right) = 
\frac{\cf}{4 \pi^2} \frac{\sqrt{\pi}}{\sqrt{8}} \ g^2(t) \ 
\left( 1 + \mathcal{O}(g^2) \right), \qquad \log \prent = 
\frac{g^2(T) \, \cf \, \me}{8 \pi T} + \mathcal{O} \left( g^4(T), 
g^4 \sqrt{t} T \right) . 
\eeq{expr1}

\subsection{Gluon condensates}
\label{cond.lo}
Insight into the flow behavior of the electric and magnetic condensates 
can be obtained by looking at their leading order expressions.  
Writing $\sqrt{t} T = x$, one gets, to $\mathcal{O} (g^2)$, \\
\begin{eqnarray}
\frac{<M>}{T^4} &=& \frac{4 g^2}{\pi^2} \ \left( \sqrt{\frac{\pi}{2}}  
\frac{1}{8 x^3} \, \Theta_3(0, e^{-8 \pi^2 x^2}) \ - \ h(x) \right) + \mathcal{O}(g^4) \nonumber \\
\frac{<E>}{T^4} &=& \frac{4 g^2}{\pi^2} \ \left( \sqrt{\frac{\pi}{2}}  
\frac{1}{16 x^3} \, \Theta_3(0, e^{-8 \pi^2 x^2}) + h(x) 
\right) + \mathcal{O}(g^4) \label{eq.lo.cond} \\
h(x) &=& \sqrt{\frac{\pi}{2}} 
\frac{2 \pi^2}{x} e^{-8 \pi^2 x^2} \, \Theta^\prime_3 (0, e^{-8 \pi^2 x^2}) 
- 8 \pi^4 \sum_{n >0} n^3 \, \Phi_c(2 \sqrt{2} \pi x n) \nonumber
\end{eqnarray}
where $\Theta_3(0, y) = \sum_n y^{n^2}$, $\Theta_3^\prime(0, y) = 
\frac{\textstyle d}{\textstyle dy} \Theta_3(0, y)$. 
From the above, 
\beq
\frac{<E+M>}{T^4} =  \frac{3 g^2}{4 \pi^2 x^3} \sqrt{\frac{\pi}{2}}  \Theta_3(0, e^{-8 \pi^2 x^2}) + \mathcal{O}(g^4).
\eeq{lo.gt}

$\Theta_3(0, e^{-8 \pi^2 x^2}) \sim \frac{\textstyle 1}{\textstyle \sqrt{8 \pi} x}$ 
as $x \to 0$. 
Using this, as $T \to 0$ \eq{lo.gt} reduces to the known leading order result
$t^2 <\mathcal{E}> = 
\lim_{T \to 0} x^4 \frac{\textstyle <E+M>}{\textstyle T^4} = 
\frac{\textstyle 3 g^2}{\textstyle 16 \pi^2}$.

It is illustrative to compare the leading order expressions, \eq{lo.cond}, with 
data. Such a comparison is shown in \fig{lo.cond}. As the figure shows, the 
differential flow behavior of $E$ and $M$ is already seen in leading order; 
however, the behavior is much more pronounced in the full theory.

\bef
\includegraphics[scale=0.7]{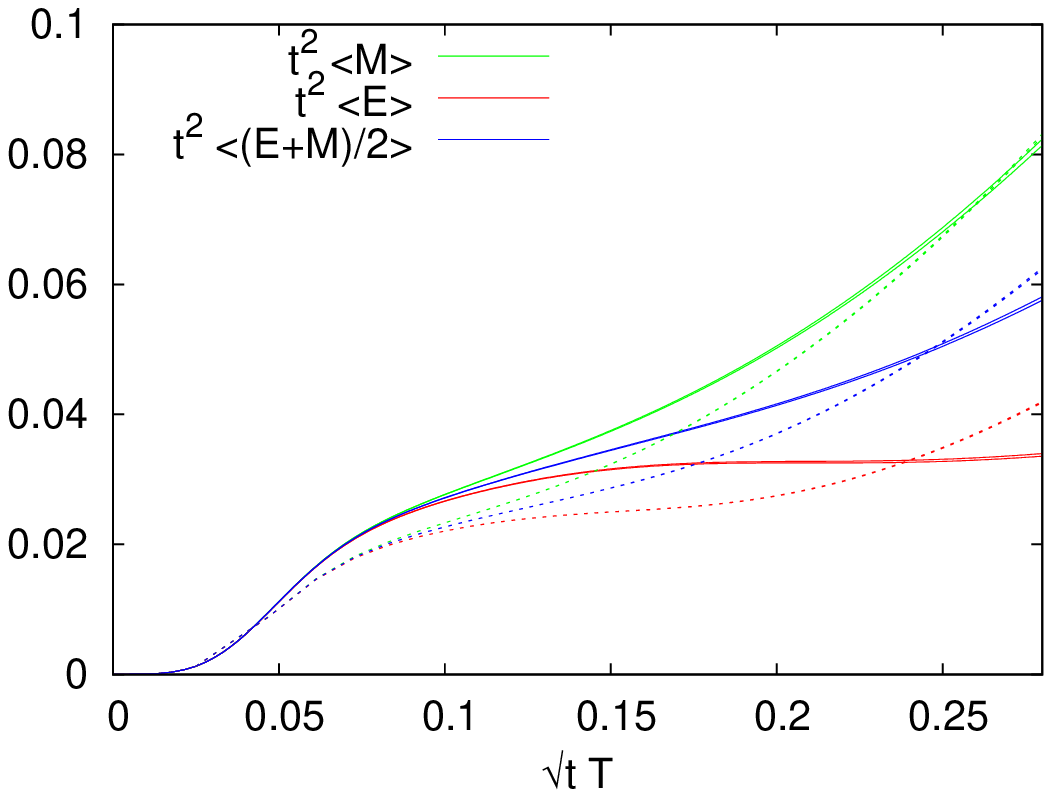}
\caption{The finite temperature condensates $t^2<M, E>$ and their average, at 
2 $T_c$, from our $N_t=8$ lattices (solid lines). Also shown, with dashed 
lines, are the corresponding leading order expressions, \eq{lo.cond}, 
\eq{lo.gt}.}
\eef{lo.cond}

Using \eq{lo.cond} and \eq{lo.gt} it is easy to calculate that
for $t \to 0$, 
\beq
\langle E(T) - E(T=0) \rangle \ =  \langle M(T) - M(T=0) \rangle \ =  
- \frac{g^2 \, (N^2 - 1) \, \pi^2}{30} \ T^4.
\eeq{lo.condsubt}

\end{document}